\documentclass[aps, twocolumn, showpacs]{revtex4}
\usepackage{graphicx}
\usepackage{dcolumn}
\usepackage{bm}
\usepackage{amsmath,amssymb}

\newcommand{\braket}[2]{\left\langle #1 \middle | #2 \right\rangle} 
\newcommand{\mele}[3]{\left\langle #1 \middle | #2 \middle | #3 \right\rangle} 
\newcommand{\ket}[1]{\left | #1 \right\rangle}

\begin{document}

\title{Fidelity susceptibilities in the one-dimensional extended Hubbard model}
\author{Wing Chi Yu}
\affiliation{Department of Physics and ITP, The Chinese University of Hong Kong, Hong
Kong, China}
\author{Shi-Jian Gu}
\altaffiliation{Email: sjgu@phy.cuhk.edu.hk}
\affiliation{Department of Physics and ITP, The Chinese University of Hong Kong, Hong
Kong, China}
\author{Hai-Qing Lin}
\affiliation{Beijing Computational Science Research Center, Beijing 100084, China}

\begin{abstract}
We investigated the fidelity susceptibility in the one-dimension
(1D) Hubbard model and the extended Hubbard model at half-filling
via the density matrix renormalization group. From the numerical
results, we argue that in the 1D Hubbard model, the fidelity
susceptibility shows a divergence at two points which is
infinitesimally close to the critical point while it is always
extensive exactly at the critical point. For the extended Hubbard
model, we found that for a properly chosen driving parameter, the
fidelity susceptibility is able to reveal the quantum phase
transitions between the PS (phase separation)-superconducting,
superconducting-CDW (charge-density wave), CDW-SDW(spin-density
wave), SDW-PS, CDW-BOW (bond-order wave), and the BOW-SDW phases.
\end{abstract}

\date{\today}
\maketitle


\section{Introduction}

At absolute zero temperature, being driven by the quantum
fluctuation rooted from the Heisenberg uncertainty principle, a
many-body system can undergo a quantum phase transition (QPT)
\cite{sachdev1999quantum} when it is tuned across some external
parameters like magnetic field. Across the transition point, the
ground state wavefunction of the system exhibits an abrupt change in
the qualitative structure. The ground state fidelity, which is a
concept from quantum information science to quantify the similarity
between two states, is expected to show a sudden drop at the
critical point. With this primary motivation, people started to
explore the role of fidelity played in QPTs
\cite{Quan2006,Zanardi2006}. The significance of the fidelity to
witness a QPT has been testified in a large variety of models
including the Dicke model \cite{Zanardi2006}, 1D XY model in a
transverse field \cite{Zanardi2006}, quadratic fermion Hamiltonians
\cite{Zanardi2007a, Cozzini2007}, and Bose-Hubbard model
\cite{Buonsante2007,Oelkers2007} (For a review, please refer to Ref.
\cite{GU2010}).

Along the streamline of fidelity, the fidelity susceptibility
\cite{You2007,Zanardi2007} was proposed. The fidelity susceptibility
is the leading response of the fidelity to the external
perturbation. It has been found to be related to the correlation
function of the driving term in the Hamiltonian \cite{You2007}.
Basing on this general relation, a scaling relation between the size
dependence of the fidelity susceptibility, the dynamic exponent, and
the scaling dimension of the driving term was also obtained through
scaling transformation \cite{CamposVenuti2007,Gu2008}. These suggest
that the fidelity susceptibility could be a potential candidate to
witness the Landau type QPTs through its divergence at the critical
point. As a pure Hilbert space geometrical quantity, the application
of the fidelity and fidelity susceptibility does not require any a
prior knowledge in the system's symmetry and this make it
advantageous to the study of QPTs. This approach has also
been successfully applied to detect the topological QPT, which fall
beyond the framework of the spontaneous symmetry-breaking theory, in
the 2D Kitaev model on a honeycomb lattice \cite{Yang2008}. Recently, a equality relating
the fidelity susceptibility and the spectral function was also derived \cite{Gu2014}. This makes the fidelity susceptibility directly measurable in experiments via neutron
scattering or the angle-resolved photoemission spectroscopy(ARPES) techniques.

However, the ability of the fidelity susceptibility to detect for
Beresinskii-Kosterlitz-Thouless (BKT) transitions
\cite{Beresinskii1971, Kosterlitz1973, Kosterlitz1974} is still
controversial. For examples, while the fidelity susceptibility was
found to be divergent at the BKT quantum critical point of the 1D
spin-half XXZ model \cite{Yang2007, Fjerestad2008, Wang2010a}, no
singularity was found when crossing the BKT transition point of the
1D asymmetric Hubbard model \cite{Gu2008}. To investigate the
ability of the fidelity susceptibility to witness a BKT transition,
in particular in the 1D Hubbard and extended Hubbard model at
half-filling, is the motivation of this work.

The fidelity susceptibility in the 1D Hubbard model has been
investigated separately by two groups \cite{You2007,
CamposVenuti2008}. You \emph{et. al.} showed that the normalized
fidelity susceptibility at the critical point is a constant even
when the system size is very large \cite{You2007}. On the other
hand, using Bosonization technique with the aid of exact
diagonalization, Venuti \emph{et. al.} argued that the normalized
fidelity susceptibility diverges in the thermodynamic limit when one
approaches the critical point on the half-filling line
\cite{CamposVenuti2008}. There is still no an agreement on whether
the fidelity susceptibility is able to signal for the BKT transition
in the model. In this work, we use the density matrix
renormalization group (DMRG) approach to calculate the fidelity
susceptibility in the model and performed a detail scaling analysis.
With the supplement of analytic calculation, we argued that the
fidelity susceptibility diverges at some infinitesimal distance away
from the critical point in the thermodynamic limit.

We also studied the fidelity susceptibility in the 1D extended
Hubbard model at half-filling. The model exhibits a very rich ground
state phase diagram consisting of the charge-density wave (CDW),
spin-density wave (SDW), phase separation (PS),
singlet-superconducting (SS), triplet-superconducting (TS) and
bond-order wave (BOW) phases. To the best of our knowledge, there is
no investigation on the fidelity susceptibility in the model
available so far. Using DMRG simulation, we investigate the ability
of the fidelity susceptibility in clarifying the ground state phase
diagram of the model. We found that the phase boundaries, except the
SS-TS one, can be roughly revealed by the fidelity susceptibility
with properly chosen driving parameters. Our finding may provide an
alternate way to study the controversies in the model's ground state
phase diagram \cite{Nakamur2000,Jeckelmann2002, Jeckelmann2003}.

The paper is organized as follows: In section \ref{sec:FS}, the
mathematical formulation of the fidelity susceptibility is reviewed.
Then the analysis of the fidelity susceptibility in the 1D Hubbard
model and the extended Hubbard model is presented in section
\ref{sec:HM} and \ref{sec:EHM} respectively. Finally, a conclusion
is given in section \ref{sec:con}.

\section{Fidelity susceptibility}\label{sec:FS}
Consider a many-body system, the Hamiltonian can be generally
written as
\begin{eqnarray}
H(\lambda)=H_0+\lambda H_I,\label{eq:H+HI}
\end{eqnarray}
where $\lambda$ is the external driving parameter and $H_I$ is the
driving Hamiltonian. Denoting the eigenstates and eigenenergies as
$\ket{\Psi_n}$ and $E_n$ respectively such that
$H(\lambda)\ket{\Psi_n(\lambda)}=E_n(\lambda)\ket{\Psi_n(\lambda)}$.
The ground state fidelity of a pure state is defined as
\cite{Zanardi2006}
\begin{eqnarray}
F(\lambda,\lambda+\delta\lambda)=\left|\braket{\Psi_0(\lambda)}{\Psi_0(\lambda+\delta\lambda)}\right|,\label{eq:F}
\end{eqnarray}
which measures the overlap between the two ground states at a
different $\lambda$. Geometrically, the fidelity represents the
angular distance between the two ground states differed by a small
value of the driving parameter $\delta\lambda$ in the Hilbert space.
It has a range between $0$ and $1$. If the two states are the same
up to a phase factor, the fidelity is equal to one. If the states
are orthogonal, the fidelity becomes zero. Across a quantum critical
point, the fidelity is expected to exhibit a significant drop as the
ground states on the two sides of the critical point has
qualitatively different structures.

For a phase transition which is not induced by the ground state
level-crossing, the ground state of the system is non-degenerated in
a finite system. Consider the system's parameter varies from $\lambda$
to $\lambda+\delta\lambda$, one can apply the non-degenerated
perturbation theory by treating $\delta\lambda H_I$ as a
perturbation. To the first order correction of the ground state, we
have
\begin{eqnarray}
\ket{\Psi_0(\lambda+\delta\lambda)}=\ket{\Psi_0(\lambda)}+\delta\lambda\sum_{n\ne
0}\frac{H_I^{n0}(\lambda)\ket{\Psi_n(\lambda)}}{E_0(\lambda)-E_n(\lambda)},\label{eq:psi1st}
\end{eqnarray}
where
\begin{eqnarray}
H_I^{n0}=\mele{\Psi_n(\lambda)}{H_I}{\Psi_0(\lambda)}.
\end{eqnarray}
Together with the normalization condition, the fidelity in Eq.
(\ref{eq:F}) to the lowest order of $\delta\lambda$ can be expressed
as
\begin{eqnarray}
F(\lambda,\lambda+\delta\lambda)\simeq
1-\frac{1}{2}(\delta\lambda)^2\sum_{n\ne
0}\frac{H_I^{n0}H_I^{0n}}{(E_0(\lambda)-E_n(\lambda))^2}.\label{eq:pert_F}
\end{eqnarray}

Recall that the maximum of the fidelity is equal to one at
$\delta\lambda=0$. It is an even function of $\delta\lambda$ and the
first order correction to the fidelity should be zero as expected.
The second order term, which is the leading response of the fidelity
to the external perturbation, is defined as the fidelity
susceptibility \cite{Zanardi2006, You2007}
\begin{eqnarray}
\chi_F(\lambda)&\equiv&\lim_{\delta\lambda\rightarrow 0}\frac{-2\ln
F}{(\delta\lambda)^2},\label{eq:fs_numeric}\\
&=&-\frac{\partial^2 F}{\partial(\delta\lambda)^2}.
\end{eqnarray}
From Eq. (\ref{eq:pert_F}), the perturbation form of the fidelity
susceptibility is given by
\begin{eqnarray}
\chi_F(\lambda)=\sum_{n\ne
0}\frac{\left|\mele{\Psi_n(\lambda)}{H_I}{\Psi_0(\lambda)}\right|^2}{(E_0(\lambda)-E_n(\lambda))^2}.
\label{eq:chiF}
\end{eqnarray}

One may realize that the form of the fidelity susceptibility is very
similar to that of the second derivative of the ground state energy
with respect to $\lambda$, i.e.
\begin{eqnarray}
\frac{\partial^2E_0(\lambda)}{\partial\lambda^2}=\sum_{n\ne
0}\frac{2\left|\mele{\Psi_n(\lambda)}{H_I}{\Psi_0(\lambda)}\right|^2}{E_0(\lambda)-E_n(\lambda)}.
\end{eqnarray}
The main difference is in the exponent of the denominator.
Therefore, one may expect that both the divergence of the fidelity
susceptibility and the second derivative of the ground state energy
at the critical point is intrinsically due to the vanishing of the
energy gap in the thermodynamic limit  \cite{Chen2008}. However, the
difference in the exponent suggests that the fidelity susceptibility
is a more sensitive seeker and may be able to detect higher order
quantum phase transitions.

\section{Analysis on the one-dimensional Hubbard
model}\label{sec:HM}

\begin{figure}[tbp]
\centering
  \includegraphics[width=8cm]{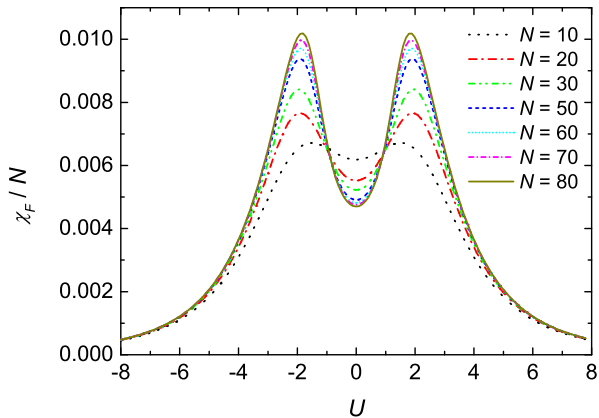}\\
  \caption{The normalized fidelity susceptibility as a function of $U$ with various system size $N$ in the 1D Hubbard model at half-filling.
  The fidelity susceptibility shows two peaks around $U=0$ while it exhibits a local minimum at $U=0$.}\label{Fig:fs_HM}
\end{figure}

The Hubbard model is the simplest model in condensed matter physics
that captures the electron-electron correlation in solids. It was
first proposed independently by Martin
Gutzwiller\cite{Gutzwiller1963}, Junjiro Kanamori\cite{Kanamori1963}
and, of course, John Hubbard\cite{Hubbard1963} in 1963, and is put
forward by Hubbard's sequential works
\cite{Hubbard1964,Hubbard1964a,Hubbard1965,Hubbard1967,Hubbard1967a}.
Originally, the model was proposed to study the itinerant
ferromagnetism in transition metals such as iron and nickel.
However, its application was later found to be far beyond this. For
examples, the model has also used in attempts to describe high-$T_c$
superconductivity. Recently with advance in cold atom experiments,
the model was also successfully realized by ultracold Fermi gas in
optical lattices \cite{Jordens2008, Schneider2008}.

The Hamiltonian of the model is given by
\begin{eqnarray}
H_{\mathrm{HM}}=-t\sum_{j=1,\sigma}\left( c_{j,\sigma }^{\dagger
}c_{j+1,\sigma }+h.c.\right)+U\sum_{j=1} n_{j,\uparrow
}n_{j,\downarrow },
\end{eqnarray}%
where $c_{j,\sigma }^{\dagger }$ and $c_{j,\sigma }$ ($\sigma
=\uparrow ,\downarrow $) is respectively the creation and
annihilation operators of an electron with a spin $\sigma $ at the
site $j$ and they satisfies the usual anti-commutation relations of
fermion operators. $n_{j,\sigma}=c_{j,\sigma }^{\dagger }c_{j,\sigma
}$ is the number operator of electrons with spin $\sigma$ at site
$j$. $t$ is the hopping amplitude of electrons between nearest
neighboring sites and is taken as one in the our analysis for
convenience. $U$ is the strength of the on-site interaction.
Originally, the $U$ term was introduced to describe the repulsive
Coulomb interaction of the electrons. One may expect only to
consider positive values of $U$. However, in some materials with
interactions like the electron-phonon or the excitonic ones, $U$ can
have an effective negative value. Without loss of generality, we
will consider the whole range of values of $U$. In the case of
half-filling, the model exhibits a BKT transition at $U=0$ and the
system transits from a metallic phase to a Mott-insulating phase as
$U$ changes from negative to positive.

Treating the $U$ term as the driving Hamiltonian, i.e.
\begin{eqnarray}
H_I=\sum_{j=1}^N n_{j,\uparrow }n_{j,\downarrow
}=\sum_{j=1}^Nc_{j\uparrow}^{\dagger}c_{j\uparrow}c_{j\downarrow}^{\dagger}c_{j\downarrow},\label{eq:HI_HM}
\end{eqnarray}
the fidelity susceptibility is calculated by DMRG simulation for
system sizes $N=10$ to $N=90$ with open boundary conditions. The
maximum number of block states kept is $m=600$. Fig. \ref{Fig:fs_HM}
shows a plot of the normalized fidelity susceptibility as a function
of the driving parameter $U$. The fidelity susceptibility is a local
minimum at the critical point $U_c=0$ and attain two maxima at some
values of $U=U_{\max}$ away from the critical point.

At the point $U=0$, the model is reduced to the tight-binding model
and analytical solution for the fidelity susceptibility can be
obtained. For open boundary conditions, the Fourier transformation
on the fermion operators is given by
\begin{eqnarray}
c_j^{\dagger}=\sqrt{\frac{2}{N+1}}\sum_k \sin(kj)c_k^{\dagger},\\
c_j=\sqrt{\frac{2}{N+1}}\sum_k \sin(kj)c_k,
\end{eqnarray}
where $k=n\pi/(N+1)$, and $n=1,2,\cdots N$. The driving Hamiltonian
in Eq. (\ref{eq:HI_HM}) becomes
\begin{eqnarray}
H_I&=&\frac{1}{2(N+1)}\sum_{k_1,k_2,k_3,k_4}\tilde{\delta}
c_{k_1,\uparrow}^{\dagger}c_{k_2,\uparrow}c_{k_3,\downarrow}^{\dagger}c_{k_4,\downarrow}.
\label{eq:HI_k}
\end{eqnarray}
where
\begin{eqnarray}
\tilde\delta&\equiv&\delta_{k_1-k_2-k_3+k_4,0}+\delta_{k_1-k_2+k_3-k_4,0}\nonumber\\
&&-\delta_{k_1-k_2-k_3-k_4,0}-\delta_{k_1-k_2+k_3+k_4,0}\nonumber\\
&&-\delta_{k_1+k_2-k_3+k_4,0}-\delta_{k_1+k_2+k_3-k_4,0}\nonumber\\
&&+\delta_{k_1+k_2-k_3-k_4,0}+\delta_{k_1+k_2+k_3+k_4,0},
\end{eqnarray}
and $\delta$ is the Kronecker delta. The ground state is given by
electrons filling up to the Fermi level and the fidelity
susceptibility is contributed by the scattering of electrons below
the Fermi level to above the Fermi level. With Eq. (\ref{eq:chiF})
and (\ref{eq:HI_k}), the fidelity susceptibility can be calculated
as
\begin{eqnarray}
\chi_F&=&\frac{1}{4(N+1)^2}\sum_{\substack{k_1,k_3>k_F \\ k_2,k_4\le
k_F}}\frac{\tilde\delta^2}{(\Delta E)^2} ,\label{eq:fs_ana_HM}
\end{eqnarray}
where $k_F$ is the Fermi wavevector and
\begin{eqnarray}
\Delta E = -2(\cos k_1+\cos k_3-\cos k_2-\cos k_4).
\end{eqnarray}
With these expressions, we can obtain the fidelity susceptibility by
carrying out the sum numerically.

\begin{figure}
\centering
  \includegraphics[width=8cm]{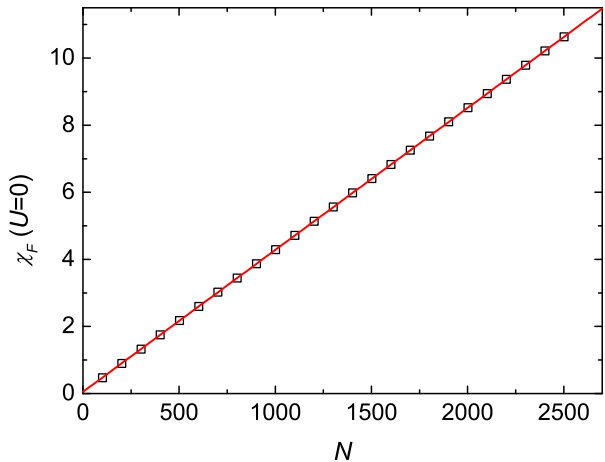}\\
  \caption{The analytical result of the fidelity susceptibility at $U=0$ in Eq. (\ref{eq:fs_ana_HM})
   as a function of $N$ in the 1D Hubbard model. The straight line
  shows a linear fitting of the data points.}
  \label{fig:ana_fs}
\end{figure}

Figure \ref{fig:ana_fs} shows a plot of the fidelity susceptibility
as a function of $N$ up to a system size of $2502$. From the figure,
one can observe that the fidelity susceptibility is linear in $N$
for a large enough system. From the slope of the straight line, we
obtained
\begin{eqnarray}
\chi_F(U=0)=0.05+0.0042N.
\end{eqnarray}
So in the thermodynamic limit, $\chi_F/N$ at $U=0$ converges to a
constant value of $0.0042\pm 10^{-7}$.

\begin{figure}[tbp]
\centering
  \includegraphics[width=8cm]{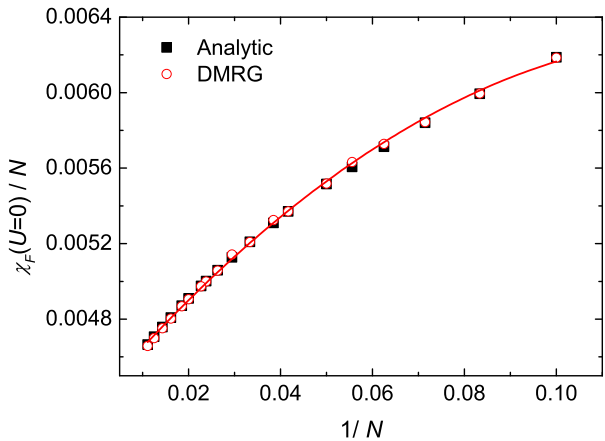}\\
  \caption{The fidelity susceptibility at $U=0$ of the 1D Hubbard model as a function of $1/N$.
  The open and solid symbol represents data obtained from DMRG and the analytical expression in Eq. (\ref{eq:fs_ana_HM}) respectively.
  The curve shows the second order polynomial fitting on the DMRG data.}\label{Fig:U0_scaling_HM}
\end{figure}

Furthermore, the normalized fidelity susceptibility at $U=0$
obtained from DMRG is compared with the analytic result in Fig.
\ref{Fig:U0_scaling_HM}. The data agrees well with each other. The
maximum discrepancy of the DMRG data from the analytic one is about
$0.5\%$. We may reasonably trust our DMRG result when doing the
analysis.

From the quadratic fitting of the DMRG data in Fig.
\ref{Fig:U0_scaling_HM}, we have
\begin{eqnarray}
\frac{\chi_F(U=0)}{N}=0.00438+0.0283\frac{1}{N}-0.104\frac{1}{N^2}.
\end{eqnarray}
In the thermodynamic limit, $\chi_F(U=0)/N\rightarrow 0.00438\pm
0.00001$, which is consistent with the analytical result to one
significant figure.


\begin{figure}[tbp]
\centering
  \includegraphics[width=8cm]{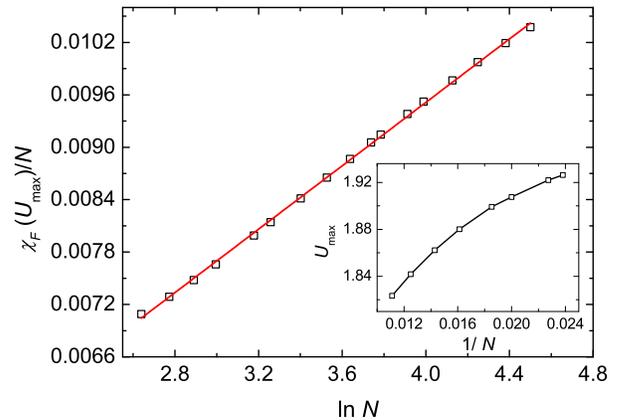}\\
  \caption{The maximum of the normalized fidelity susceptibility of the 1D Hubbard model as a function of $\ln N$. The straight
  line shows the linear fitting of the data points. Inset shows a plot of $U_{\max}$ at which the maximum of fidelity susceptibility occurs as a function of $1/N$.}\label{Fig:scaling_HM}
\end{figure}

Next, let's investigate the scaling behavior of the two peaks away
from the critical point. They are symmetric about the $U=0$ axis. So
we can only pick one of the peaks to do the finite size scaling
analysis. The maximum value of the normalized fidelity
susceptibility is plotted as a function of $\ln N$ in Fig.
\ref{Fig:scaling_HM}. The data points falls on a straight line. From
linear fitting, we have
\begin{eqnarray}
\frac{\chi_F(U_{\max})}{N}\sim\ln N.
\end{eqnarray}
In the thermodynamic limit, the normalized fidelity susceptibility
shows a logarithmic divergence.


The inset of figure \ref{Fig:scaling_HM} shows the scaling behavior
of the value of $U$ at which the maximum of the fidelity
susceptibility takes place. We see that as the system size
increases, $U_{\max}$ decreases. In the thermodynamic limit, we
expect it to tend to a
point which is infinitesimally close to zero. 

Integrating the above analyses, we argue that in the thermodynamic
limit, the fidelity susceptibility is expected to show a logarithmic
divergence at two points $0^+$ and $0^-$ which is infinitesimally
close to the true critical point $U=0$. Exactly at the critical
point, the normalized fidelity susceptibility is always a local
minimum and have a constant value of  $0.004$. We try to understand
this phenomena by noting that the ground state of the system is a
charge-density wave (CDW) and a spin-density wave (SDW) on the sides
of negative and positive $U$ respectively. However, at the critical
point $U=0$, the system is a free electron system which is
completely different from a CDW or a SDW. This point may be
considered as another phase. So as $U$ is tuned from $-\infty$ to
$+\infty$, the ground state of the system undergoes two abrupt
changes, one from CDW to a free electrons and the other time from
free electrons to SDW. This gives rise to the two peaks in the
fidelity susceptibility. For the fidelity susceptibility to be
continuous in finite systems, it must experience a local minimum in
between. As the two peaks have to be symmetric as a result of
particle-hole symmetry in the model, the local minimum has
to occur at $U=0$.
%


\section{Analysis on the one-dimensional extended Hubbard
model}\label{sec:EHM}

\begin{figure}
  \includegraphics[width=8cm]{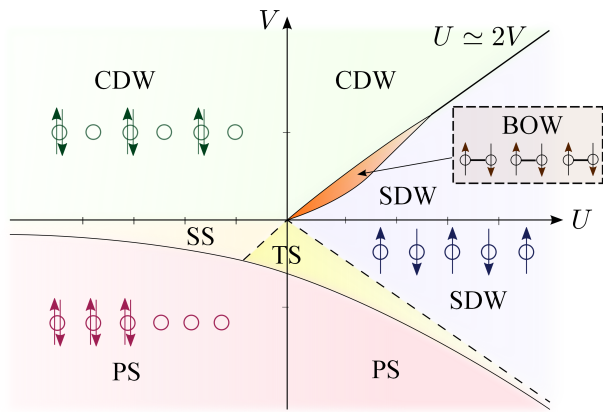}\\
  \caption{A schematic drawing of the ground state phase diagram of the 1D extended Hubbard model at half-filling.}\label{Fig:EHM_phase}
\end{figure}

Despite the historical importance of the Hubbard model, it is
limited to describe materials in which long-ranged Coulomb
interaction plays an essential role. Examples included organic
conductors such as TTF-TCNQ \cite{Farges1994, Claessen2002} and
$\pi$-conjugated polymers like polydiacetylene \cite{Baeriswyl1992}.
To model these materials, it is necessary to invoke at least the
nearest-neighbor Coulomb interaction. This leads to the extended
Hubbard model described by the Hamiltonian
\begin{eqnarray}
H_{\mathrm{EHM}}&=&-t\sum_{j=1,\sigma}\left( c_{j,\sigma
}^{\dagger }c_{j+1,\sigma }+h.c.\right) \nonumber\\
&&+U\sum_{j=1} n_{j,\uparrow }n_{j,\downarrow }+V\sum_{j=1} n_j
n_{j+1}, \label{eq:H_EHM}
\end{eqnarray}
where $n_j=n_{j,\uparrow}+n_{j,\downarrow}$. The $V$ term in the
Hamiltonian captures the nearest-neighbor interaction and it
contributes whenever two neighboring sites are simultaneously
occupied.

In the past few decades, a variety of technique was implanted to
study the ground state phase diagram of the model. It was found that
the model exhibits a very rich phase diagram (See Fig.
\ref{Fig:EHM_phase}). In the strong coupling limit ($|U|, |V|>> t$),
perturbation analysis suggested the existence of charge-density wave
(CDW), the spin-density wave (SDW) and phase separation (PS) phases
in the ground state phase diagram \cite{Emery1976,
Fowler1978, Dongen1994}. 
In the weak coupling limit ($|U|, |V|<< t$), analytic studied
through the g-ology \cite{Emery1979, Solyom1979, Fourcade1984} and
bosonization method \cite{Cannon1990, Voit1992} in the past also
gave some insight on the ground state phase diagram of the model .
By investigating various correlation functions in a given region,
these theories predicted the existence of the CDW, the SDW, singlet
superconducting (SS) and the triplet superconducting (TS) phase.

Although much effort has been devoted to study the extended Hubbard
model in the past, there remain controversies in its ground state
phase diagram. For positive $U$ and $V$, while the strong coupling
theories predicted the CDW-SDW phase transition to be first order,
the weak coupling theories tell that it is continuous. By studying
the excitation spectra with numerical exact diagonalization,
Nakamura pointed out that there also exist a spontaneous dimerized
phase, the bond-order wave (BOW) phase, in a narrow region between
the SDW phase and the CDW phase up to a tricritical point
\cite{Nakamur2000}. On the other hand, Jeckelmann argued that the
BOW phase only exits on a short segment of the critical line, rather
than a stripe, of the CDW-SDW transition from DMRG studies
\cite{Jeckelmann2002, Jeckelmann2003}. Subsequent efforts using
quantum Monte Carlo stimulation \cite{Sengupta2002, Sandvik2004},
density matrix renormalization group \cite{Zhang2004, Ejima2007},
and analytical methods \cite{Tsuchiizu2002, Tsuchiizu2004} have been
devoted to clarify this issue in the ground state phase diagram.
Recently, the concept from quantum information science, namely the
entanglement entropy, was also used to explore the problem
\cite{Mund2009, Liu2011}. Although most of the studies confirmed the
existence of the BOW phase, the shape of it and the position of the
tricritical point still have not settled into agreement.

In the following, let's take $t=1$ again for convenience and
consider the case of half-filling. Unless otherwise specified, the
data presented are obtained from DMRG with open boundary conditions.
The fidelity susceptibility depends on the path of the driving
Hamiltonian. In our analysis, we considered taking the $U$ term and
the $V$ term as the driving Hamiltonian respectively. Note that one
can also consider the case of varying both $U$ and $V$ together, but
we will not discuss it here.

\subsection{$U$ as the driving parameter}

\begin{figure}[tbp]
\centering
  \includegraphics[width=6.5cm]{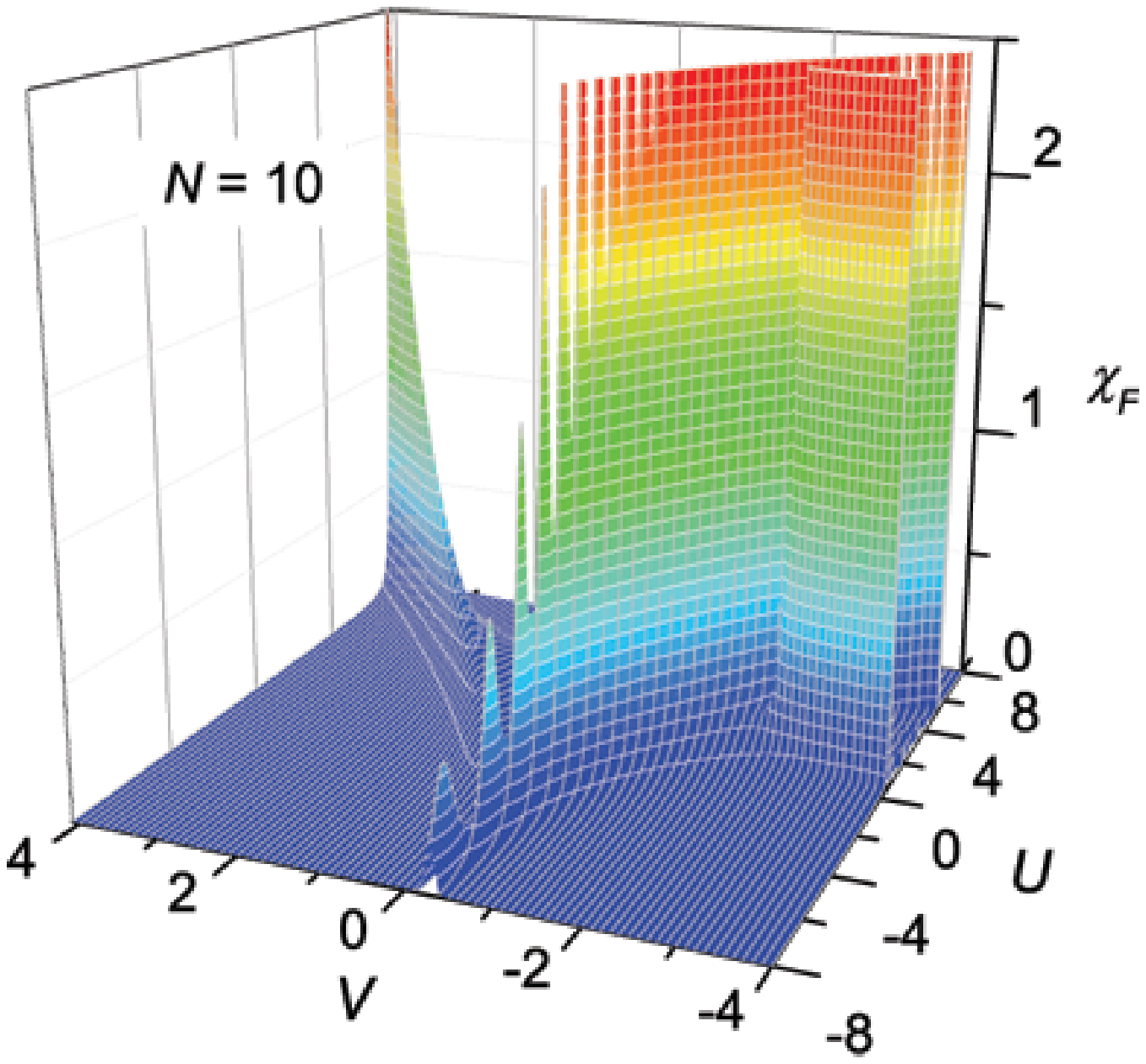}\\
  \includegraphics[width=8cm]{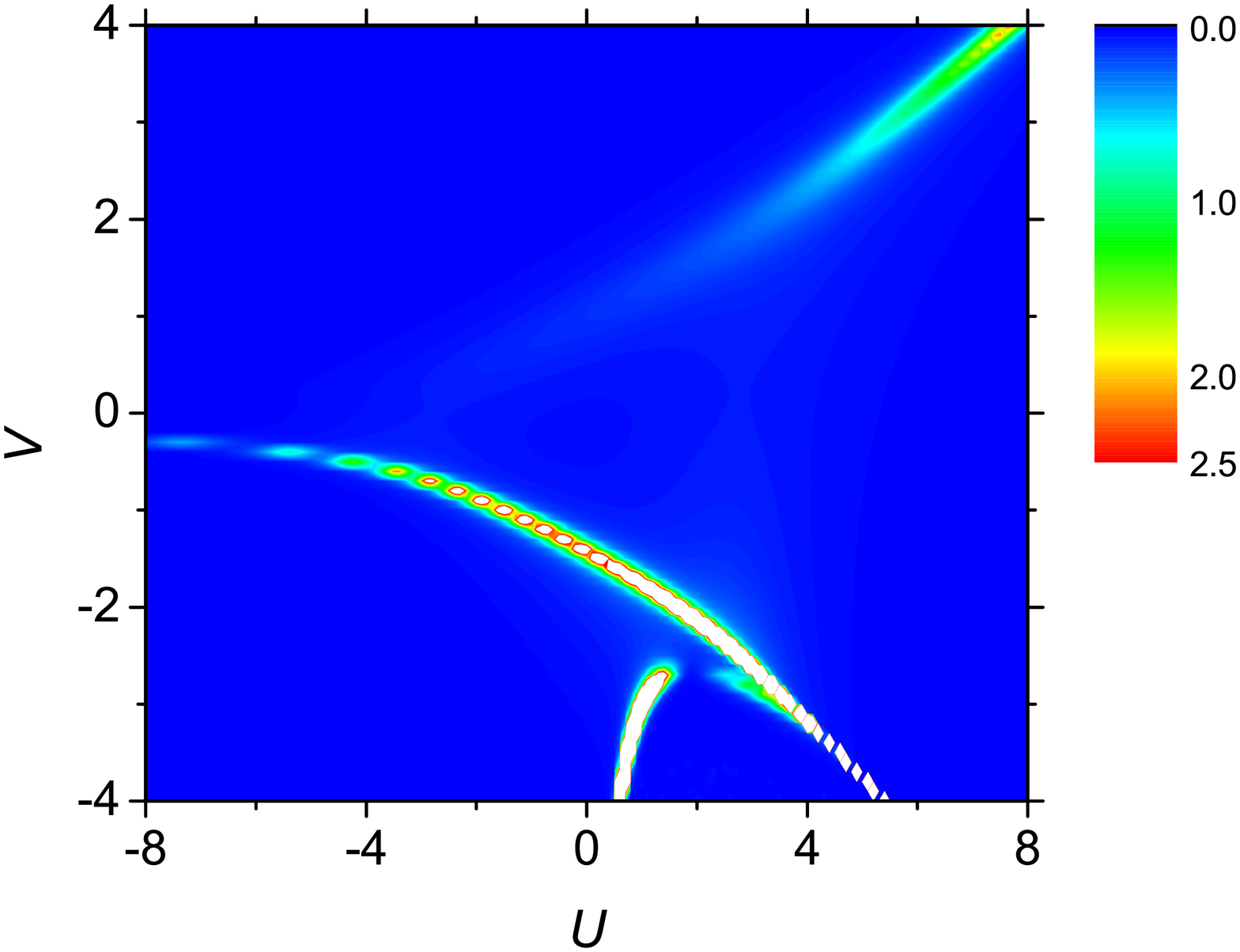}\\
  \caption{(Top) A 3D plot of the fidelity susceptibility
  as a function of $U$ and $V$ in the 1D extended Hubbard model.
  (Bottom) A contour of the 3D plot on the $U$-$V$ plane.
  $U$ is treated as the driving parameter in this case. The data is
  obtained from numerical exact diagonalization for $N=10$ with
  periodic boundary conditions.}\label{Fig:fs_EHM_ED_dU}
\end{figure}

The driving Hamiltonian is taken as
\begin{eqnarray}
H_I=\sum_{j=1}^N n_{j,\uparrow }n_{j,\downarrow }.
\end{eqnarray}
To have a brief picture of the overall phase diagram, we first
carried out numerical exact diagonalization to calculate the
fidelity susceptibility for a small system, i.e. $N=10$. To minimize
the size effect, we used periodic boundary conditions in the
calculation.

Figure \ref{Fig:fs_EHM_ED_dU} shows a 3D plot of the fidelity
susceptibility as a function of $U$ and $V$, and also the contour on
the $U$-$V$ plane. From the figure, we can observe that, even with
such a small system size, the fidelity susceptibility has captured
the phase boundaries of the CDW-SDW, SDW-PS, PS-CDW transitions. In
the negative value of $V$, the fidelity susceptibility shows a very
sharp peak at the phase boundary between the PS and SDW phase. The
huge peak there may be understood by the fact that the transition is
a first order transition. The transition is caused by a
level-crossing between the ground state and the excited state in the
energy spectrum. The crossing occurs even in a very small system. So
the fidelity as a measure of the overlap between the ground state
differed by a small $U$ should have a sharp drop at the critical
point. This drop in fidelity is then reflected in the peak of the
fidelity susceptibility.

Moreover, in the contour plot, there is also another sharp boundary
in the PS phase in the positive $U$ and negative $V$ regime (The
boundary going through the point $U\simeq 0.6$ and $V=-4$). However,
this is not a true critical line. It is just a finite size effect.
Consider a PS state
\begin{eqnarray}
\text{PS(a):} \hspace{10pt}
\uparrow\downarrow\hspace{6pt}\uparrow\downarrow\hspace{6pt}\uparrow\downarrow\hspace{6pt}\uparrow\downarrow\hspace{6pt}\uparrow\downarrow\hspace{6pt}
0\hspace{6pt}0\hspace{6pt}0\hspace{6pt}0\hspace{6pt}0,
\end{eqnarray}
having energy $(2N-4)V+NU/2$. The ground state in the $U$,
$V\rightarrow -\infty$ limit is a superposition of the above state
and its translational symmetric states. Now if $U$ increases and
gradually becomes positive (but still not reaching the SDW phase),
the on-site interaction term would like to break the electron pairs
and one doubly occupied site tends to be singly occupied. The weight
of
\begin{eqnarray}
\text{PS(b):} \hspace{10pt}0\hspace{6pt}0\hspace{6pt}
\uparrow\hspace{6pt}\uparrow\downarrow\hspace{6pt}\uparrow\downarrow\hspace{6pt}\uparrow\downarrow\hspace{6pt}\uparrow\downarrow\hspace{6pt}
\downarrow\hspace{6pt}0\hspace{6pt}0\hspace{6pt}
\end{eqnarray}
increases and at some positive value of $U$, PS(b) would become the
dominate configuration in the ground state. The peaks in the
fidelity susceptibility is in fact correspond to the crossover
between the two PS states. Note that the configuration in PS(b) has
an energy of $(2N-4)V+(N/2-1)U$. The energy difference between the
PS(a) and PS(b) configuration is $U$ and this difference becomes
negligible when $N$ is large. So the crossover peak in the fidelity
susceptibility would be suppressed in the thermodynamic limit.

For the case of positive $U$ and $V$, the contour in Fig.
\ref{Fig:fs_EHM_ED_dU} also reveals the SDW-CDW phase transition.
The peaks in the fidelity susceptibility are stronger in the large
$U$ and $V$ regime than in the small $U$ and $V$ region. This is
because the phase transition there is a discontinuous one. For the
weak and intermediate coupling regime, the phase boundary and its
detail, for examples the BOW phase, is merely resolved. The
transition here is believed to be continuous for the CDW-BOW and BKT
for the BOW-SDW transition. It would thus expected to be harder to
realize in a small system. To have a more detail clarification of
the phase diagram, we calculated the fidelity susceptibility for a
larger system size using DMRG. Two particular paths, varying U while
fixing $V=1$ and $V=-0.5$ respectively, will be discussed in the
following.

\subsubsection{\emph{Case I: V=1}}

\begin{figure}[tbp]
\centering
  \includegraphics[width=8cm]{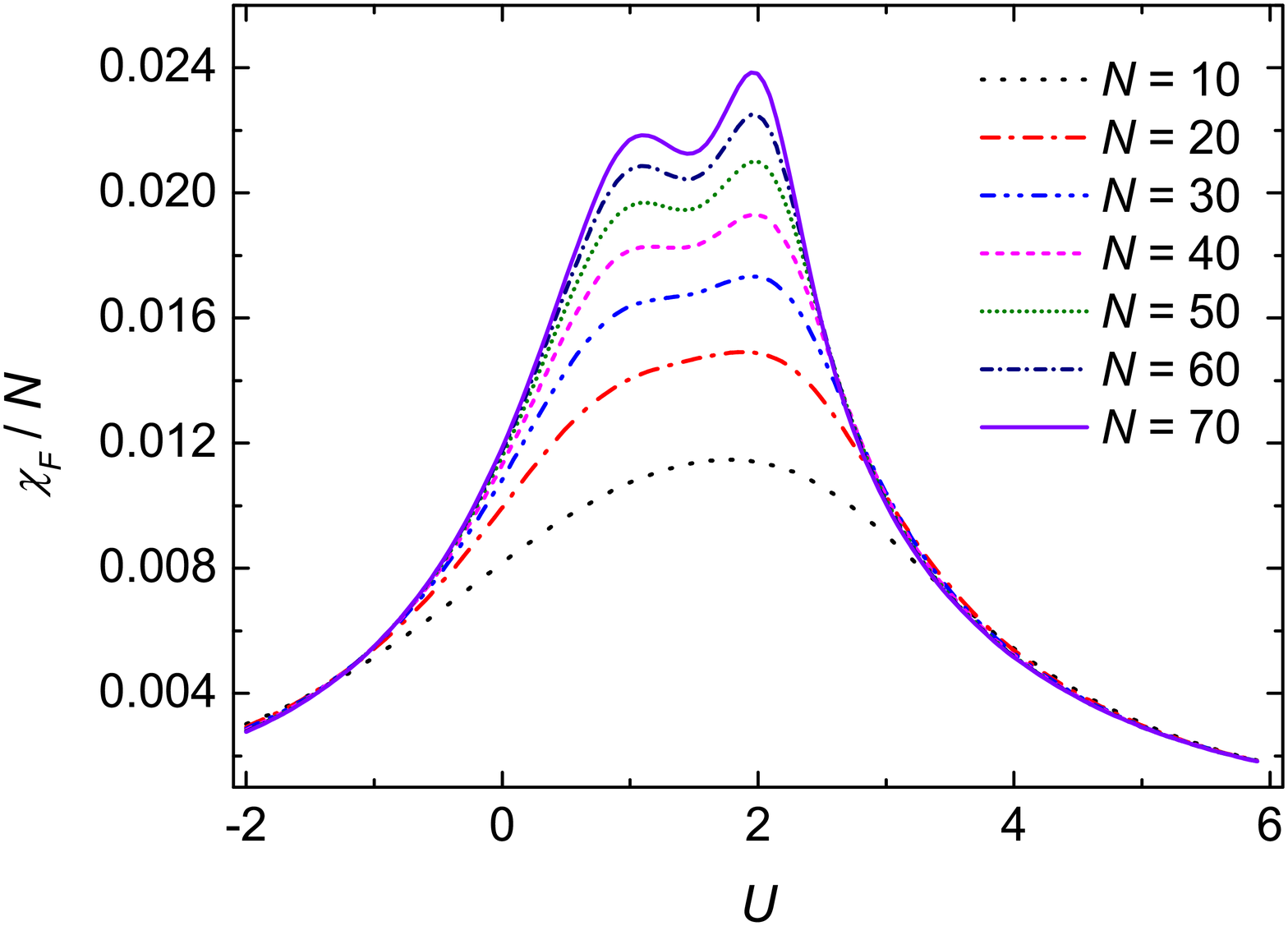}\\
  \caption{The normalized fidelity susceptibility as a function of $U$ in the 1D extended Hubbard model at half-filling for $V=1$.}\label{Fig:fs_EHM_V1}
\end{figure}

Figure \ref{Fig:fs_EHM_V1} shows a plot of the normalized fidelity
susceptibility as a function of the driving parameter $U$. For
system size smaller than $40$, there is only one maximum occurs
around $U=2$. However, as the system size increases, there is
another local maximum build up around $U=1$. While $\chi_F/N$ is
intensive away from the critical region, the two local maxima in the
vicinity of the critical point become larger and larger as the
system gets bigger. We can reasonably suspect that the maximum
around $U_{\max1}\simeq 1$ and $U_{\max2}\simeq 2$ is corresponding
to the SDW-BOW and BOW-CDW transition respectively.
%


\begin{figure}[tbp]
\centering
  \includegraphics[width=8cm]{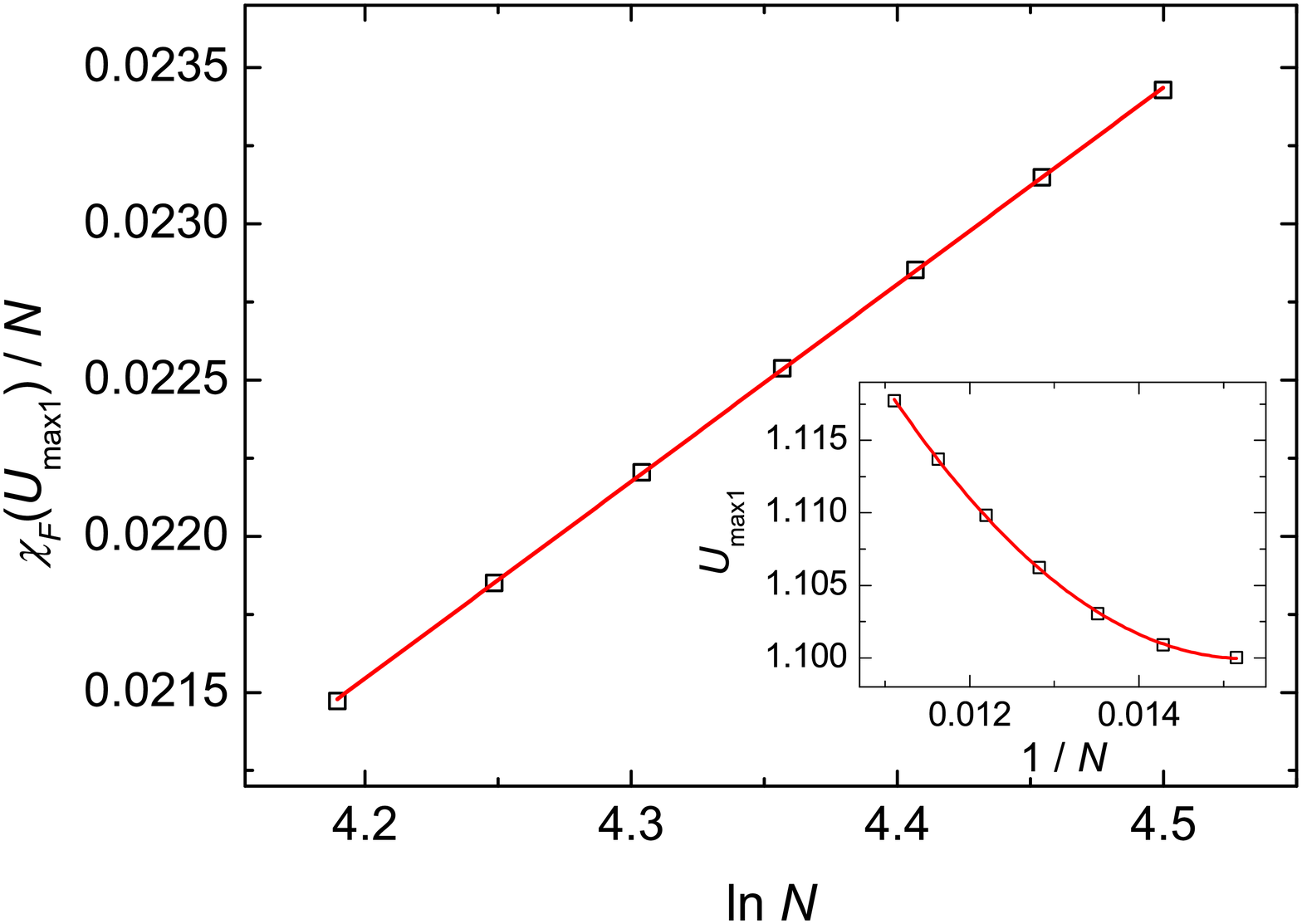}\\
  \caption{A semi-ln plot of the maximum of the fidelity susceptibility around $U_{\max1}\simeq1$
  as a function of the system size in the 1D extended Hubbard model at
  half-filling. Here $V=1$ and $U$ is taken as the driving parameter. The straight line shows the linear fitting of the data
  point. The inset shows a plot of $U_{\max1}$ as a function of $1/N$ and the curve shows a second order polynomial fitting of the data.}
  \label{Fig:scaling_EHM_V1}
\end{figure}


In Fig. \ref{Fig:scaling_EHM_V1}, the maximum of the fidelity
susceptibility at $U_{\max1}$ is plotted against $\ln N$. The data
points fall perfectly onto a straight line. We have
\begin{eqnarray}
\frac{\chi_F(U_{\max1})}{N}\sim \ln N.
\end{eqnarray}

The inset of figure \ref{Fig:scaling_EHM_V1} shows a plot of the
value of $U_{\max1}$ as a function of $1/N$. The data points are
fitted by a second order polynomial curve. From the fitting, we
found that
\begin{eqnarray}
\lim_{N\rightarrow\infty}U_{\max1}= 1.340\pm 0.004.\label{eq:Umax1}
\end{eqnarray}


%

\begin{figure}[t]
\centering
  \includegraphics[width=8cm]{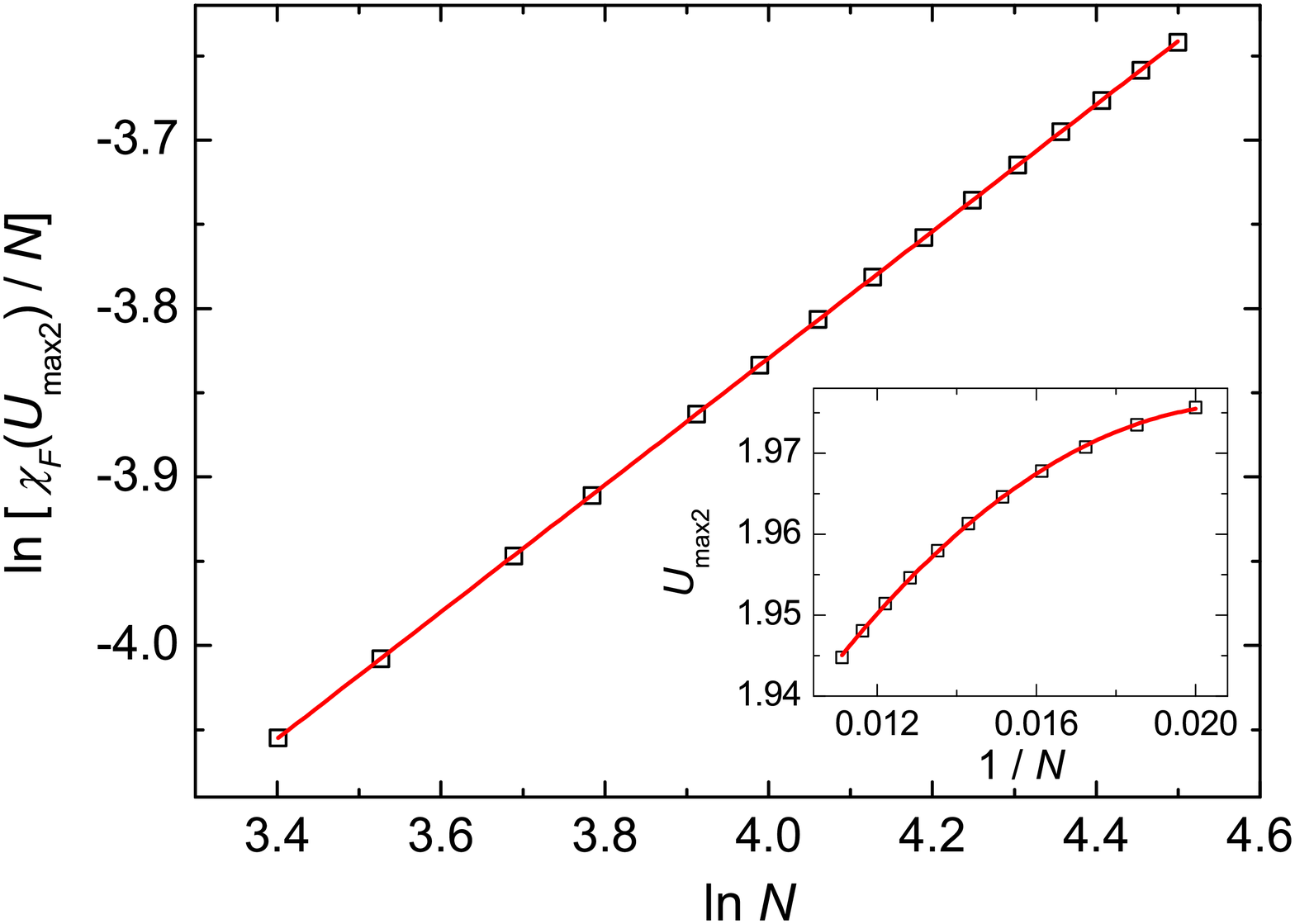}\\
  \caption{A ln-ln plot of the maximum of the fidelity susceptibility around $U=2$
  as a function of the system size in the 1D extended Hubbard model at
  half-filling. Here $V=1$ and $U$ is taken as the driving parameter. The straight line, which shows the linear fitting of the data point,
  has a slope of $0.3784\pm0.0001$. Inset shows a plot of $U_{\max2}$ as a function of $1/N$.
  The curve shows a second order polynomial fitting of the data points.}\label{Fig:scaling_EHM_V1_max2}
\end{figure}

For the maximum at $U_{\max2}$, the scaling behavior of the fidelity
susceptibility is shown in Fig. \ref{Fig:scaling_EHM_V1_max2}. The
normalized fidelity susceptibility as a function of $N$ is plotted
in a natural logarithmic scale in the figure. We can see that the
normalized fidelity susceptibility scales algebraically with the
system size. From linear fitting, we obtained
\begin{eqnarray}
\frac{\chi_F(U_{\max2})}{N}\sim N^{0.3784\pm0.0001}.
\end{eqnarray}

Moreover, the value of $U_{\max2}$ is plotted as a function of $1/N$
in the inset of Fig. \ref{Fig:scaling_EHM_V1_max2}. As the system
size increases, $U_{\max2}$ tends to decrease. The data points are
well fitted onto
a quadratic curve. 
We obtained, in the thermodynamic limit,
\begin{eqnarray}
\lim_{N\rightarrow\infty}U_{\max2}=1.842\pm 0.002.\label{eq:Umax2}
\end{eqnarray}

From the above analysis, we argue that the fidelity susceptibility
diverges at two values of $U$, which indeed corresponds to the
SDW-BOW and BOW-CDW transition respectively. While in the former
case the divergence is logarithmic, the fidelity susceptibility
diverges algebraically in the latter case. In the thermodynamic
limit, the two critical points tends to a different value in Eq.
(\ref{eq:Umax1}) and (\ref{eq:Umax2}). So instead of just a line
segment, we tend to believe that the BOW phase is a stripe in the
ground state phase diagram on the $U-V$ plane. At $V=1$, the width
of the BOW phase is found to be
\begin{eqnarray}
\Delta U_{\text{BOW}}=0.502\pm 0.006.
\end{eqnarray}
This width obtained agrees roughly with the analytical result
obtained from g-ology in Ref. \cite{Tsuchiizu2002}. However, the
critical points obtained in Eq. (\ref{eq:Umax1}) and
(\ref{eq:Umax2}) deviate from those obtained by calculating the
correlation functions \cite{Zhang2004} or the spin and the charge
gaps in other studies \cite{Sandvik2004, Ejima2007}. We are not sure
the apparent agreement in $\Delta U_{\text{BOW}}$ with the g-ology
result is just a coincidence or there are more physics to be
explored.

\subsubsection{\emph{Case II: V=-0.5}}

\begin{figure}[tbp]
\centering
  \includegraphics[width=8cm]{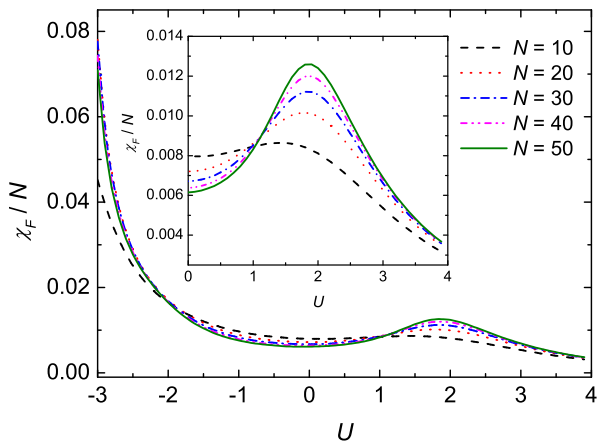}\\
  \caption{The fidelity susceptibility as a function of $U$ in the 1D extend Hubbarded model at half-filling for $V=-0.5$.
  The inset shows a close-up of the fidelity susceptibility around $U=0$ to $U=4$.}\label{Fig:fs_EHM_V05}
\end{figure}

Next, let's consider the path along $V=-0.5$. In this case, the
system goes through the PS, SS, TS, and the SDW phase as $U$
increases. Fig. \ref{Fig:fs_EHM_V05} shows a plot of the fidelity
susceptibility as a function of $U$ for various system sizes. Here
we have only shown the result for $U\gtrsim -3$. For the data with
$U$ smaller than this value, the fidelity susceptibility is strongly
fluctuating with large amplitudes and becomes unreliable. This is a
result of symmetry breaking in the simulation. In principle, the
ground state is given by a superposition of PS(a) and its
translational symmetric states for a finite system. However, the
potential barrier between these translational invariant state for
PS(a) state is very large especially in a large system. If we start
with a random initial wavefunction in the stimulation, the system
would converge to one of the translational invariant states and
leads to a strongly fluctuating fidelity susceptibility.
Nevertheless, though the critical point cannot be exactly located,
we can still notice the PS-SS phase transition when the fidelity
susceptibility changes from a fluctuating pattern into a smooth
function.



\begin{figure}[tbp]
\centering
  \includegraphics[width=8cm]{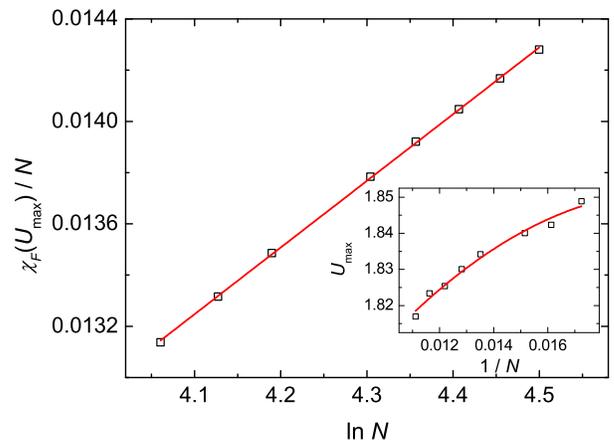}\\
  \caption{A semi-ln plot of the maximum of the fidelity susceptibility
  as a function of the system size in the 1D extended Hubbard model at
  half-filling. Here $V=-0.5$ and $U$ is taken as the driving parameter. The straight line is a linear fitting of the data points. Inset shows a plot of $U_{\max}$ as a function of $1/N$ in the 1D extended Hubbard model at half-filling for $V=-0.5$.
  The curve shows a second order polynomial fitting of the data points.}\label{Fig:scaling_EHM_V05}
\end{figure}

In Fig. \ref{Fig:fs_EHM_V05}, one can also observe a small peak in
the fidelity susceptibility around $U\simeq 2$. This peak indeed
indicate the transition from the superconducting phase to the SDW
phase of the model. From the inset, we can see that $\chi_F/N$ shows
a non-trivial size dependence. This local maximum of the normalized
fidelity susceptibility is plotted as a function of $\ln N$ in Fig.
\ref{Fig:scaling_EHM_V05}. The data points agrees well with the
linear fitted line. So we have
\begin{eqnarray}
\frac{\chi_F(U_{\max})}{N}\sim\ln N,
\end{eqnarray}
and it diverges in the thermodynamic limit.

To determine the location of $U_{\max}$ in the thermodynamic limit,
we plotted the value of it as a function of $1/N$ in the inset of
Fig. \ref{Fig:scaling_EHM_V05}. From the second order polynomial
fitting, we have obtained
\begin{eqnarray}
\lim_{N\rightarrow\infty}U_{\max}=1.70\pm 0.03.
\end{eqnarray}

For the SS-TS transition, we were not able to observe any
significant peaks in the fidelity susceptibility around the
transition point. This may due to the limitation in the small system
sizes being stimulated. Another difficulty may arise from the narrow
width of the superconducting phase for $V=-0.5$ in the phase diagram. Together
with the effect of the strong peak from PS-SS transition, the SS-TS
transition peak may be suppressed.


\subsection{$V$ as the driving parameter}

\begin{figure}[tbp]
\centering
  \includegraphics[width=7cm]{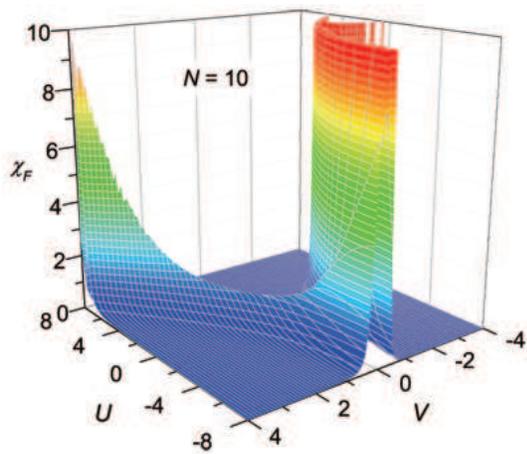}\\
  \includegraphics[width=8cm]{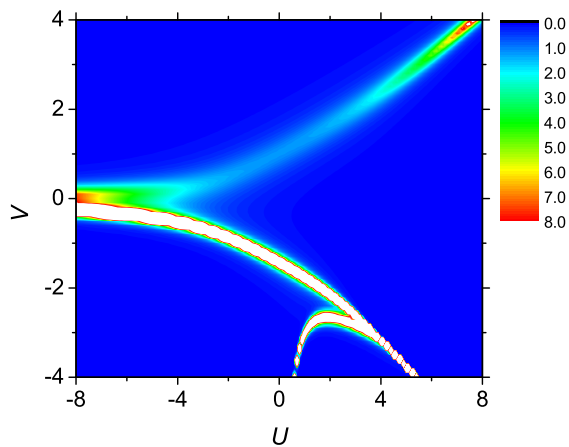}\\
  \caption{(Top) A 3D plot of the fidelity susceptibility
  as a function of $U$ and $V$ in the 1D extended Hubbard model.
  (Bottom) A contour of the 3D plot on the $U$-$V$ plane.
  $V$ is treated as the driving parameter in this case. The data is
  obtained from numerical exact diagonalization for $N=10$ with
  periodic boundary conditions.}\label{Fig:fs_EHM_ED_dV}
\end{figure}

Next we consider the $V$ term as the driving Hamiltonian, i.e.
\begin{eqnarray}
H_I=\sum_{j=1}^{N-1} n_j n_{j+1}.
\end{eqnarray}
Using numerical exact diagonalization with periodic boundary
conditions, we calculated the fidelity susceptibility for a system
of $10$ sites. The result is shown in Fig. \ref{Fig:fs_EHM_ED_dV} as
a function of $U$ and $V$. From the contour, one can clearly see the
transition line for the PS-SDW, PS-CDW, CDW-SDW transitions, and the
cross-over line in the PS phase. The fidelity susceptibility even
shows a stronger peak around these boundaries than that in the
$U$-driven case. As expected, it is a path-dependent quantity and we
may also argue that the V-driven one is a more sensitive seeker to
the continuous and the discontinuous phase transition  in the
extended Hubbard model than the V-driven one.

To have a finer structure of the phase boundaries, we chose two
particular paths to study the fidelity susceptibility. The first one
is along the $U=2$ line. The system goes through the PS, TS, SDW,
BOW, and CDW phases when $V$ increases from a very negative value.
The other path is along the $U=-2$ line. In this case, the system
goes through the PS, SS, and CDW phases as $V$ increases.

\subsubsection{\emph{Case I: $U=2$}}

\begin{figure}[tbp]
\centering
  \includegraphics[width=8cm]{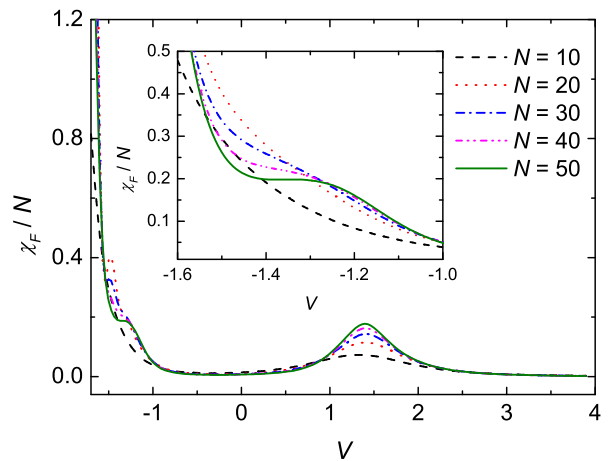}\\
  \caption{A plot of the normalized fidelity susceptibility as a function of $V$ in the 1D extended Hubbard model.
  Here $U=2$. The inset shows a close-up of the fidelity susceptibility around $V=-1.3$.}\label{Fig:fs_EHM_U2}
\end{figure}

In Fig. \ref{Fig:fs_EHM_U2}, the fidelity susceptibility is plotted
as a function of $V$ for various system sizes. For $V$ smaller than
the range of value shown, the fidelity susceptibility is fluctuating
with strong amplitudes. The change from the fluctuating pattern into
a smooth one indicates roughly the boundary of the PS phase. Around
the region of $V=-1.3$, the fidelity susceptibility has a point of
inflection for $N\geq 30$. We expect that somewhere in this region,
the fidelity susceptibility will develop a peak for large enough
system and thus indicates a TS-SDW phase transition. However, we are
not able to observe this with our current computational power.

%
%

\begin{figure}[tbp]
\centering
  \includegraphics[width=8cm]{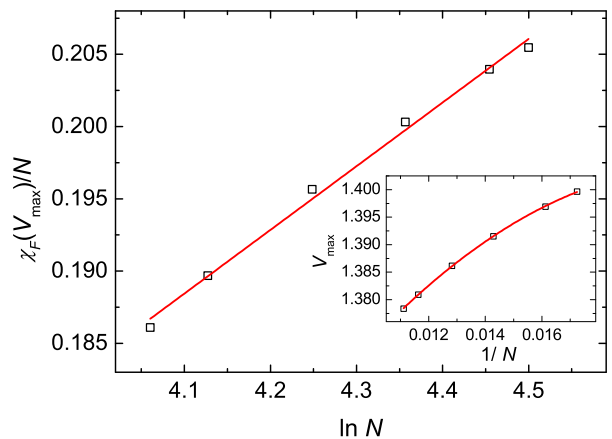}\\
  \caption{A plot of the maximum of the fidelity susceptibility around $V=1.5$ as a function of $\ln N$ in the 1D extended Hubbard model.
  Here $U=2$. The straight line shows the linear fitting of the data points. The inset shows the scaling behavior of $V_{\max}$ as a function of $1/N$.
  The curve is a second order polynomial fitting of the data.}\label{Fig:scaling_EHM_U2}
\end{figure}

Around $V=1.5$, the fidelity susceptibility show a local maximum
with non-trivial size dependence. As the system size increases, the
amplitude of this local maximum also increases. From Fig.
\ref{Fig:scaling_EHM_U2}, we found that
\begin{eqnarray}
\frac{\chi_F(V_{\max})}{N}\sim\ln N.
\end{eqnarray}
The nearest-neighbor interaction strength at which the local maximum
take place, .i.e. $V_{\max}$, is also plotted as a function of $1/N$
in the inset of Fig. \ref{Fig:scaling_EHM_U2}. From the second order
polynomial fitting, we obtained
\begin{eqnarray}
\lim_{N\rightarrow\infty} V_{\max}=1.296\pm0.002.
\end{eqnarray}
In the thermodynamic limit, the fidelity susceptibility diverges and
signals for a SDW-CDW phase transition.

However, for the transition into the BOW phase, we does not observe
any significant signature in the fidelity susceptibility in this
case.

\subsubsection{\emph{Case II: $U=-2$}}

\begin{figure}[tbp]
\centering
  \includegraphics[width=8cm]{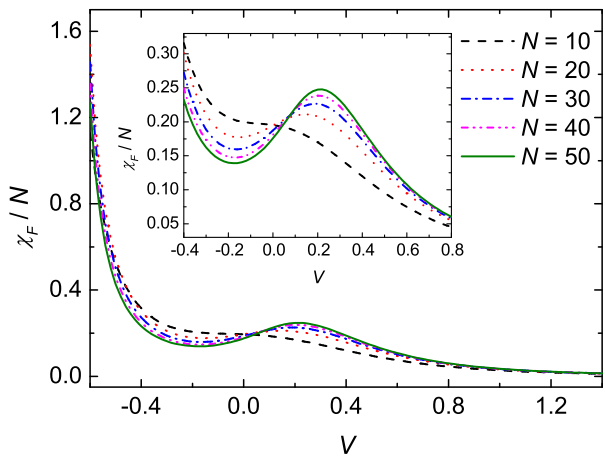}\\
  \caption{A plot of the fidelity susceptibility as a function of $V$ at $U=-2$ in the 1D extended Hubbard model.
  The inset shows a close-up around $V=0.2$.}\label{Fig:fs_EHM_U-2}
\end{figure}

\begin{figure}[tbp]
\centering
  \includegraphics[width=8cm]{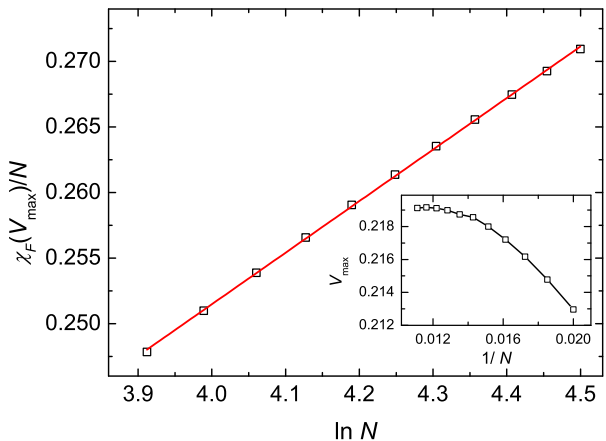}\\
  \caption{A semi-ln plot of the maximum of fidelity susceptibility around $V=0.2$ versus $N$ in the 1D extended Hubbard model. Here $U=-2$.
  The straight line shows a linear fitting of the data points. The inset shows a plot of $V_{\max}$ as a function of $1/N$.}\label{Fig:scaling_EHM_U-2}
\end{figure}

Figure \ref{Fig:fs_EHM_U-2} shows a plot of $\chi_F/N$ as a function
of $V$ for $U=-2$ in the system for various $N$. Again in the
vicinity of the PS-SS phase transition around $V=-0.6$, the
amplitude of the fidelity susceptibility shows a different pattern
on the PS side to the SS side. Around $V=0.2$, the fidelity
susceptibility exhibits a local maximum. The magnitude of the
maximum increases as the system size increases. From the semi-ln
plot in Fig. \ref{Fig:scaling_EHM_U-2}, we found that
\begin{eqnarray}
\frac{\chi_F(V_{\max})}{N}\sim\ln N,
\end{eqnarray}
and it diverges logarithmically in the thermodynamic limit. This
indicates a phase transition between the SS and CDW phase.

%

In the inset of Fig. \ref{Fig:scaling_EHM_U-2}, the $V_{\max}$
around $V=0.2$ is shown as a function of $1/N$. The location of
$V_{\max}$ increases initially with the system size, but the rate of
increase slows down as the system size increases. When reaching the
maximum stimulated size $N=90$, the value of $V_{\max}$ shows a
small drop compare with the one for $N=86$. For the system size
stimulated, we highly suspect that we still have not entered the
scaling region and we would expect $V_{\max}$ to decrease further as
we increase the system size. A satisfactory scaling analysis on
$V_{\max}$ is not possible with our current computational power in
this case.

\section{Conclusion} \label{sec:con}

To conclude, we have investigated the fidelity susceptibility in the
1D Hubbard model and the extended Hubbard model. In the Hubbard
model, we argued that the fidelity susceptibility diverges at two
points infinitesimally close to the critical point while it remains
extensive exactly at the critical point. 

For the extended Hubbard model, we studied the fidelity
susceptibility for the U-driven and V-driven cases. From the
divergence of the fidelity susceptibility, one may notice the
existence of the transition between PS-superconducting,
superconducting-CDW, CDW-SDW, SDW-PS phases. Moreover, for the
U-driven case, evidence for the CDW-BOW and BOW-SDW phase transition
along $V=1$ was also found. However, in most of the above case, the
critical points of the phase transition obtained from the scaling
analysis showed discrepancies from previous studies (For examples in
Ref. \cite{Zhang2004}, \cite{Sandvik2004}, \cite{Ejima2007}). One of
the reasons for the discrepancy may due to the limitation of
system size being simulated. Another reason may due to the behavior
of the fidelity susceptibility in an order-to-order phase
transition. Like in the Hubbard model, the divergence in the
fidelity susceptibility may not occurs exactly at the critical
point. Similar founding was also obtained in the 1D Bose-Hubbard
model in recent studies \cite{Carrasquilla2013, Lacki2014}. The
physical reason behind still require further studies.

W. C. Yu thank Wen-Long You for helpful discussion.
This work is supported by the Earmarked Grant Research from the
Research Grants Council of HKSAR, China (Project No. CUHK 401212).


\begin{thebibliography}{55}
\expandafter\ifx\csname
natexlab\endcsname\relax\def\natexlab#1{#1}\fi
\expandafter\ifx\csname bibnamefont\endcsname\relax
  \def\bibnamefont#1{#1}\fi
\expandafter\ifx\csname bibfnamefont\endcsname\relax
  \def\bibfnamefont#1{#1}\fi
\expandafter\ifx\csname citenamefont\endcsname\relax
  \def\citenamefont#1{#1}\fi
\expandafter\ifx\csname url\endcsname\relax
  \def\url#1{\texttt{#1}}\fi
\expandafter\ifx\csname urlprefix\endcsname\relax\def\urlprefix{URL
}\fi \providecommand{\bibinfo}[2]{#2}
\providecommand{\eprint}[2][]{\url{#2}}

\bibitem[{\citenamefont{Sachdev}(1999)}]{sachdev1999quantum}
\bibinfo{author}{\bibfnamefont{S.}~\bibnamefont{Sachdev}},
  \emph{\bibinfo{title}{Quantum phase transitions}}
  (\bibinfo{publisher}{Cambridge University Press, Cambridge, England},
  \bibinfo{year}{1999}).

\bibitem[{\citenamefont{Quan et~al.}(2006)\citenamefont{Quan, Song, Liu,
  Zanardi, and Sun}}]{Quan2006}
\bibinfo{author}{\bibfnamefont{H.~T.} \bibnamefont{Quan}},
  \bibinfo{author}{\bibfnamefont{Z.}~\bibnamefont{Song}},
  \bibinfo{author}{\bibfnamefont{X.~F.} \bibnamefont{Liu}},
  \bibinfo{author}{\bibfnamefont{P.}~\bibnamefont{Zanardi}}, \bibnamefont{and}
  \bibinfo{author}{\bibfnamefont{C.}~\bibnamefont{Sun}},
  \bibinfo{journal}{Phys. Rev. Lett.} \textbf{\bibinfo{volume}{96}},
  \bibinfo{pages}{140604} (\bibinfo{year}{2006}).

\bibitem[{\citenamefont{Zanardi and Paunkovi\ifmmode~\acute{c}\else
  \'{c}\fi{}}(2006)}]{Zanardi2006}
\bibinfo{author}{\bibfnamefont{P.}~\bibnamefont{Zanardi}} \bibnamefont{and}
  \bibinfo{author}{\bibfnamefont{N.}~\bibnamefont{Paunkovi\ifmmode~\acute{c}\else
  \'{c}\fi{}}}, \bibinfo{journal}{Phys. Rev. E} \textbf{\bibinfo{volume}{74}},
  \bibinfo{pages}{031123} (\bibinfo{year}{2006}).

\bibitem[{\citenamefont{Zanardi
  et~al.}(2007{\natexlab{a}})\citenamefont{Zanardi, Cozzini, and
  Giorda}}]{Zanardi2007a}
\bibinfo{author}{\bibfnamefont{P.}~\bibnamefont{Zanardi}},
  \bibinfo{author}{\bibfnamefont{M.}~\bibnamefont{Cozzini}}, \bibnamefont{and}
  \bibinfo{author}{\bibfnamefont{P.}~\bibnamefont{Giorda}},
  \bibinfo{journal}{J. Stat. Mech.: Theory Exp.}
  \textbf{\bibinfo{volume}{2007}}, \bibinfo{pages}{L02002}
  (\bibinfo{year}{2007}{\natexlab{a}}).

\bibitem[{\citenamefont{Cozzini et~al.}(2007)\citenamefont{Cozzini, Giorda, and
  Zanardi}}]{Cozzini2007}
\bibinfo{author}{\bibfnamefont{M.}~\bibnamefont{Cozzini}},
  \bibinfo{author}{\bibfnamefont{P.}~\bibnamefont{Giorda}}, \bibnamefont{and}
  \bibinfo{author}{\bibfnamefont{P.}~\bibnamefont{Zanardi}},
  \bibinfo{journal}{Phys. Rev. B} \textbf{\bibinfo{volume}{75}},
  \bibinfo{pages}{014439} (\bibinfo{year}{2007}).

\bibitem[{\citenamefont{Buonsante and Vezzani}(2007)}]{Buonsante2007}
\bibinfo{author}{\bibfnamefont{P.}~\bibnamefont{Buonsante}} \bibnamefont{and}
  \bibinfo{author}{\bibfnamefont{A.}~\bibnamefont{Vezzani}},
  \bibinfo{journal}{Phys. Rev. Lett.} \textbf{\bibinfo{volume}{98}},
  \bibinfo{pages}{110601} (\bibinfo{year}{2007}).

\bibitem[{\citenamefont{Oelkers and Links}(2007)}]{Oelkers2007}
\bibinfo{author}{\bibfnamefont{N.}~\bibnamefont{Oelkers}} \bibnamefont{and}
  \bibinfo{author}{\bibfnamefont{J.}~\bibnamefont{Links}},
  \bibinfo{journal}{Phys. Rev. B} \textbf{\bibinfo{volume}{75}},
  \bibinfo{pages}{115119} (\bibinfo{year}{2007}).

\bibitem[{\citenamefont{Gu}(2010)}]{GU2010}
\bibinfo{author}{\bibfnamefont{S.-J.} \bibnamefont{Gu}}, \bibinfo{journal}{Int.
  J. Mod. Phys. B} \textbf{\bibinfo{volume}{24}}, \bibinfo{pages}{4371}
  (\bibinfo{year}{2010}).

\bibitem[{\citenamefont{You et~al.}(2007)\citenamefont{You, Li, and
  Gu}}]{You2007}
\bibinfo{author}{\bibfnamefont{W.-L.} \bibnamefont{You}},
  \bibinfo{author}{\bibfnamefont{Y.-W.} \bibnamefont{Li}}, \bibnamefont{and}
  \bibinfo{author}{\bibfnamefont{S.-J.} \bibnamefont{Gu}},
  \bibinfo{journal}{Phys. Rev. E} \textbf{\bibinfo{volume}{76}},
  \bibinfo{pages}{022101} (\bibinfo{year}{2007}).

\bibitem[{\citenamefont{Zanardi
  et~al.}(2007{\natexlab{b}})\citenamefont{Zanardi, Giorda, and
  Cozzini}}]{Zanardi2007}
\bibinfo{author}{\bibfnamefont{P.}~\bibnamefont{Zanardi}},
  \bibinfo{author}{\bibfnamefont{P.}~\bibnamefont{Giorda}}, \bibnamefont{and}
  \bibinfo{author}{\bibfnamefont{M.}~\bibnamefont{Cozzini}},
  \bibinfo{journal}{Phys. Rev. Lett.} \textbf{\bibinfo{volume}{99}},
  \bibinfo{pages}{100603} (\bibinfo{year}{2007}{\natexlab{b}}).

\bibitem[{\citenamefont{Campos~Venuti and Zanardi}(2007)}]{CamposVenuti2007}
\bibinfo{author}{\bibfnamefont{L.}~\bibnamefont{Campos~Venuti}}
  \bibnamefont{and} \bibinfo{author}{\bibfnamefont{P.}~\bibnamefont{Zanardi}},
  \bibinfo{journal}{Phys. Rev. Lett.} \textbf{\bibinfo{volume}{99}},
  \bibinfo{pages}{095701} (\bibinfo{year}{2007}).

\bibitem[{\citenamefont{Gu et~al.}(2008)\citenamefont{Gu, Kwok, Ning, and
  Lin}}]{Gu2008}
\bibinfo{author}{\bibfnamefont{S.-J.} \bibnamefont{Gu}},
  \bibinfo{author}{\bibfnamefont{H.-M.} \bibnamefont{Kwok}},
  \bibinfo{author}{\bibfnamefont{W.-Q.} \bibnamefont{Ning}}, \bibnamefont{and}
  \bibinfo{author}{\bibfnamefont{H.-Q.} \bibnamefont{Lin}},
  \bibinfo{journal}{Phys. Rev. B} \textbf{\bibinfo{volume}{77}},
  \bibinfo{pages}{245109} (\bibinfo{year}{2008}).

\bibitem[{\citenamefont{Yang et~al.}(2008)\citenamefont{Yang, Gu, Sun, and
  Lin}}]{Yang2008}
\bibinfo{author}{\bibfnamefont{S.}~\bibnamefont{Yang}},
  \bibinfo{author}{\bibfnamefont{S.-J.} \bibnamefont{Gu}},
  \bibinfo{author}{\bibfnamefont{C.-P.} \bibnamefont{Sun}}, \bibnamefont{and}
  \bibinfo{author}{\bibfnamefont{H.-Q.} \bibnamefont{Lin}},
  \bibinfo{journal}{Phys. Rev. A} \textbf{\bibinfo{volume}{78}},
  \bibinfo{pages}{012304} (\bibinfo{year}{2008}).

\bibitem{Gu2014} S.-J. Gu and W. C. Yu, arXiv:1408.2199 (2014).

\bibitem[{\citenamefont{Beresinskii}(1971)}]{Beresinskii1971}
\bibinfo{author}{\bibfnamefont{V.~L.} \bibnamefont{Beresinskii}},
  \bibinfo{journal}{Sov. Phys. JETP} \textbf{\bibinfo{volume}{32}},
  \bibinfo{pages}{493} (\bibinfo{year}{1971}).

\bibitem[{\citenamefont{Kosterlitz and Thouless}(1973)}]{Kosterlitz1973}
\bibinfo{author}{\bibfnamefont{J.~M.} \bibnamefont{Kosterlitz}}
  \bibnamefont{and} \bibinfo{author}{\bibfnamefont{D.~J.}
  \bibnamefont{Thouless}}, \bibinfo{journal}{J. Phys. C}
  \textbf{\bibinfo{volume}{6}}, \bibinfo{pages}{1181} (\bibinfo{year}{1973}).

\bibitem[{\citenamefont{Kosterlitz}(1974)}]{Kosterlitz1974}
\bibinfo{author}{\bibfnamefont{J.~M.} \bibnamefont{Kosterlitz}},
  \bibinfo{journal}{J. Phys. C} \textbf{\bibinfo{volume}{7}},
  \bibinfo{pages}{1046} (\bibinfo{year}{1974}).

\bibitem[{\citenamefont{Yang}(2007)}]{Yang2007}
\bibinfo{author}{\bibfnamefont{M.-F.} \bibnamefont{Yang}},
  \bibinfo{journal}{Phys. Rev. B} \textbf{\bibinfo{volume}{76}},
  \bibinfo{pages}{180403} (\bibinfo{year}{2007}).

\bibitem[{\citenamefont{Fj{\ae}restad}(2008)}]{Fjerestad2008}
\bibinfo{author}{\bibfnamefont{J.~O.} \bibnamefont{Fj{\ae}restad}},
  \bibinfo{journal}{J. Stat. Mech.: Theory Exp.}
  \textbf{\bibinfo{volume}{2008}}, \bibinfo{pages}{P07011}
  (\bibinfo{year}{2008}).

\bibitem[{\citenamefont{Wang et~al.}(2010)\citenamefont{Wang, Feng, and
  Chen}}]{Wang2010a}
\bibinfo{author}{\bibfnamefont{B.}~\bibnamefont{Wang}},
  \bibinfo{author}{\bibfnamefont{M.}~\bibnamefont{Feng}}, \bibnamefont{and}
  \bibinfo{author}{\bibfnamefont{Z.-Q.} \bibnamefont{Chen}},
  \bibinfo{journal}{Phys. Rev. A} \textbf{\bibinfo{volume}{81}},
  \bibinfo{pages}{064301} (\bibinfo{year}{2010}).

\bibitem[{\citenamefont{Campos~Venuti et~al.}(2008)\citenamefont{Campos~Venuti,
  Cozzini, Buonsante, Massel, Bray-Ali, and Zanardi}}]{CamposVenuti2008}
\bibinfo{author}{\bibfnamefont{L.}~\bibnamefont{Campos~Venuti}},
  \bibinfo{author}{\bibfnamefont{M.}~\bibnamefont{Cozzini}},
  \bibinfo{author}{\bibfnamefont{P.}~\bibnamefont{Buonsante}},
  \bibinfo{author}{\bibfnamefont{F.}~\bibnamefont{Massel}},
  \bibinfo{author}{\bibfnamefont{N.}~\bibnamefont{Bray-Ali}}, \bibnamefont{and}
  \bibinfo{author}{\bibfnamefont{P.}~\bibnamefont{Zanardi}},
  \bibinfo{journal}{Phys. Rev. B} \textbf{\bibinfo{volume}{78}},
  \bibinfo{pages}{115410} (\bibinfo{year}{2008}).

\bibitem[{\citenamefont{Nakamura}(2000)}]{Nakamur2000}
\bibinfo{author}{\bibfnamefont{M.}~\bibnamefont{Nakamura}},
  \bibinfo{journal}{Phys. Rev. B} \textbf{\bibinfo{volume}{61}},
  \bibinfo{pages}{16377}
  (\bibinfo{year}{2000}).

\bibitem[{\citenamefont{Jeckelmann}(2002)}]{Jeckelmann2002}
\bibinfo{author}{\bibfnamefont{E.}~\bibnamefont{Jeckelmann}},
  \bibinfo{journal}{Phys. Rev. Lett.} \textbf{\bibinfo{volume}{89}},
  \bibinfo{pages}{236401}(\bibinfo{year}{2002}).

\bibitem[{\citenamefont{Jeckelmann}(2003)}]{Jeckelmann2003}
\bibinfo{author}{\bibfnamefont{E.}~\bibnamefont{Jeckelmann}},
  \bibinfo{journal}{Phys. Rev. Lett.} \textbf{\bibinfo{volume}{91}},
  \bibinfo{pages}{089702}
  (\bibinfo{year}{2003}).

\bibitem[{\citenamefont{Chen et~al.}(2008)\citenamefont{Chen, Wang, Hao, and
  Wang}}]{Chen2008}
\bibinfo{author}{\bibfnamefont{S.}~\bibnamefont{Chen}},
  \bibinfo{author}{\bibfnamefont{L.}~\bibnamefont{Wang}},
  \bibinfo{author}{\bibfnamefont{Y.}~\bibnamefont{Hao}}, \bibnamefont{and}
  \bibinfo{author}{\bibfnamefont{Y.}~\bibnamefont{Wang}},
  \bibinfo{journal}{Phys. Rev. A} \textbf{\bibinfo{volume}{77}},
  \bibinfo{pages}{032111} (\bibinfo{year}{2008}).

\bibitem[{\citenamefont{Gutzwiller}(1963)}]{Gutzwiller1963}
\bibinfo{author}{\bibfnamefont{M.~C.} \bibnamefont{Gutzwiller}},
  \bibinfo{journal}{Phys. Rev. Lett.} \textbf{\bibinfo{volume}{10}},
  \bibinfo{pages}{159} (\bibinfo{year}{1963}).

\bibitem[{\citenamefont{Kanamori}(1963)}]{Kanamori1963}
\bibinfo{author}{\bibfnamefont{J.}~\bibnamefont{Kanamori}},
  \bibinfo{journal}{Prog. Theor. Phys.} \textbf{\bibinfo{volume}{30}},
  \bibinfo{pages}{275} (\bibinfo{year}{1963}).

\bibitem[{\citenamefont{Hubbard}(1963)}]{Hubbard1963}
\bibinfo{author}{\bibfnamefont{J.}~\bibnamefont{Hubbard}},
  \bibinfo{journal}{Proc. R. Soc. A} \textbf{\bibinfo{volume}{276}},
  \bibinfo{pages}{238} (\bibinfo{year}{1963}).

\bibitem[{\citenamefont{Hubbard}(1964{\natexlab{a}})}]{Hubbard1964}
\bibinfo{author}{\bibfnamefont{J.}~\bibnamefont{Hubbard}},
  \bibinfo{journal}{Proc. R. Soc. A} \textbf{\bibinfo{volume}{277}},
  \bibinfo{pages}{237} (\bibinfo{year}{1964}{\natexlab{a}}).

\bibitem[{\citenamefont{Hubbard}(1964{\natexlab{b}})}]{Hubbard1964a}
\bibinfo{author}{\bibfnamefont{J.}~\bibnamefont{Hubbard}},
  \bibinfo{journal}{Proc. R. Soc. A} \textbf{\bibinfo{volume}{281}},
  \bibinfo{pages}{401} (\bibinfo{year}{1964}{\natexlab{b}}).

\bibitem[{\citenamefont{Hubbard}(1965)}]{Hubbard1965}
\bibinfo{author}{\bibfnamefont{J.}~\bibnamefont{Hubbard}},
  \bibinfo{journal}{Proc. R. Soc. A} \textbf{\bibinfo{volume}{285}},
  \bibinfo{pages}{542} (\bibinfo{year}{1965}).

\bibitem[{\citenamefont{Hubbard}(1967{\natexlab{a}})}]{Hubbard1967}
\bibinfo{author}{\bibfnamefont{J.}~\bibnamefont{Hubbard}},
  \bibinfo{journal}{Proc. R. Soc. A} \textbf{\bibinfo{volume}{296}},
  \bibinfo{pages}{100} (\bibinfo{year}{1967}{\natexlab{a}}).

\bibitem[{\citenamefont{Hubbard}(1967{\natexlab{b}})}]{Hubbard1967a}
\bibinfo{author}{\bibfnamefont{J.}~\bibnamefont{Hubbard}},
  \bibinfo{journal}{Proc. R. Soc. A} \textbf{\bibinfo{volume}{296}},
  \bibinfo{pages}{82} (\bibinfo{year}{1967}{\natexlab{b}}).

\bibitem[{\citenamefont{J{\"o}rdens et~al.}(2008)\citenamefont{J{\"o}rdens,
  Strohmaier, G{\"u}nter, Moritz, and Esslinger}}]{Jordens2008}
\bibinfo{author}{\bibfnamefont{R.}~\bibnamefont{J{\"o}rdens}},
  \bibinfo{author}{\bibfnamefont{N.}~\bibnamefont{Strohmaier}},
  \bibinfo{author}{\bibfnamefont{K.}~\bibnamefont{G{\"u}nter}},
  \bibinfo{author}{\bibfnamefont{H.}~\bibnamefont{Moritz}}, \bibnamefont{and}
  \bibinfo{author}{\bibfnamefont{T.}~\bibnamefont{Esslinger}},
  \bibinfo{journal}{Nature} \textbf{\bibinfo{volume}{455}},
  \bibinfo{pages}{204} (\bibinfo{year}{2008}).

\bibitem[{\citenamefont{Schneider et~al.}(2008)\citenamefont{Schneider,
  Hackerm{\"u}ller, Will, Best, Bloch, Costi, Helmes, Rasch, and
  Rosch}}]{Schneider2008}
\bibinfo{author}{\bibfnamefont{U.}~\bibnamefont{Schneider}},
  \bibinfo{author}{\bibfnamefont{L.}~\bibnamefont{Hackerm{\"u}ller}},
  \bibinfo{author}{\bibfnamefont{S.}~\bibnamefont{Will}},
  \bibinfo{author}{\bibfnamefont{T.}~\bibnamefont{Best}},
  \bibinfo{author}{\bibfnamefont{I.}~\bibnamefont{Bloch}},
  \bibinfo{author}{\bibfnamefont{T.}~\bibnamefont{Costi}},
  \bibinfo{author}{\bibfnamefont{R.}~\bibnamefont{Helmes}},
  \bibinfo{author}{\bibfnamefont{D.}~\bibnamefont{Rasch}}, \bibnamefont{and}
  \bibinfo{author}{\bibfnamefont{A.}~\bibnamefont{Rosch}},
  \bibinfo{journal}{Science} \textbf{\bibinfo{volume}{322}},
  \bibinfo{pages}{1520} (\bibinfo{year}{2008}).

\bibitem[{\citenamefont{Farges}(1994)}]{Farges1994}
\bibinfo{author}{\bibfnamefont{J.~P.} \bibnamefont{Farges}},
  \emph{\bibinfo{title}{Organic conductors}} (\bibinfo{publisher}{Dekker},
  \bibinfo{year}{1994}).

\bibitem[{\citenamefont{Claessen et~al.}(2002)\citenamefont{Claessen, Sing,
  Schwingenschl\"ogl, Blaha, Dressel, and Jacobsen}}]{Claessen2002}
\bibinfo{author}{\bibfnamefont{R.}~\bibnamefont{Claessen}},
  \bibinfo{author}{\bibfnamefont{M.}~\bibnamefont{Sing}},
  \bibinfo{author}{\bibfnamefont{U.}~\bibnamefont{Schwingenschl\"ogl}},
  \bibinfo{author}{\bibfnamefont{P.}~\bibnamefont{Blaha}},
  \bibinfo{author}{\bibfnamefont{M.}~\bibnamefont{Dressel}}, \bibnamefont{and}
  \bibinfo{author}{\bibfnamefont{C.~S.} \bibnamefont{Jacobsen}},
  \bibinfo{journal}{Phys. Rev. Lett.} \textbf{\bibinfo{volume}{88}},
  \bibinfo{pages}{096402} (\bibinfo{year}{2002}).

\bibitem[{\citenamefont{Baeriswyl and Kiess}(1992)}]{Baeriswyl1992}
\bibinfo{author}{\bibfnamefont{D.}~\bibnamefont{Baeriswyl}} \bibnamefont{and}
  \bibinfo{author}{\bibfnamefont{H.}~\bibnamefont{Kiess}},
  \emph{\bibinfo{title}{Conjugated conducting polymers}}
  (\bibinfo{publisher}{Springer}, \bibinfo{year}{1992}).

\bibitem[{\citenamefont{Emery}(1976)}]{Emery1976}
\bibinfo{author}{\bibfnamefont{V.~J.} \bibnamefont{Emery}},
  \bibinfo{journal}{Phys. Rev. B} \textbf{\bibinfo{volume}{14}},
  \bibinfo{pages}{2989} (\bibinfo{year}{1976}).

\bibitem[{\citenamefont{Fowler}(1978)}]{Fowler1978}
\bibinfo{author}{\bibfnamefont{M.}~\bibnamefont{Fowler}},
  \bibinfo{journal}{Phys. Rev. B} \textbf{\bibinfo{volume}{17}},
  \bibinfo{pages}{2989} (\bibinfo{year}{1978}).

\bibitem[{\citenamefont{van Dongen}(1994)}]{Dongen1994}
\bibinfo{author}{\bibfnamefont{P.~G.~J.} \bibnamefont{van Dongen}},
  \bibinfo{journal}{Phys. Rev. B} \textbf{\bibinfo{volume}{49}},
  \bibinfo{pages}{7904} (\bibinfo{year}{1994}).

\bibitem[{\citenamefont{Emery}(1979)}]{Emery1979}
\bibinfo{author}{\bibfnamefont{V.}~\bibnamefont{Emery}},
  \emph{\bibinfo{title}{Highly conducting one-dimensional solids}}
  (\bibinfo{publisher}{Plenum, New York}, \bibinfo{year}{1979}).

\bibitem[{\citenamefont{S{\'o}lyom}(1979)}]{Solyom1979}
\bibinfo{author}{\bibfnamefont{J.}~\bibnamefont{S{\'o}lyom}},
  \bibinfo{journal}{Adv. Phys.} \textbf{\bibinfo{volume}{28}},
  \bibinfo{pages}{201}
  (\bibinfo{year}{1979}).

\bibitem[{\citenamefont{Fourcade and Sproken}(1984)}]{Fourcade1984}
\bibinfo{author}{\bibfnamefont{B.}~\bibnamefont{Fourcade}} \bibnamefont{and}
  \bibinfo{author}{\bibfnamefont{G.}~\bibnamefont{Sproken}},
  \bibinfo{journal}{Phys. Rev. B} \textbf{\bibinfo{volume}{29}},
  \bibinfo{pages}{5089} (\bibinfo{year}{1984}).

\bibitem[{\citenamefont{Cannon and Fradkin}(1990)}]{Cannon1990}
\bibinfo{author}{\bibfnamefont{J.}~\bibnamefont{Cannon}} \bibnamefont{and}
  \bibinfo{author}{\bibfnamefont{E.}~\bibnamefont{Fradkin}},
  \bibinfo{journal}{Phys. Rev. B} \textbf{\bibinfo{volume}{41}},
  \bibinfo{pages}{9435} (\bibinfo{year}{1990}).

\bibitem[{\citenamefont{Voit}(1992)}]{Voit1992}
\bibinfo{author}{\bibfnamefont{J.}~\bibnamefont{Voit}}, \bibinfo{journal}{Phys.
  Rev. B} \textbf{\bibinfo{volume}{45}}, \bibinfo{pages}{4027}
  (\bibinfo{year}{1992}).

\bibitem[{\citenamefont{Sengupta et~al.}(2002)\citenamefont{Sengupta, Sandvik,
  and Campbell}}]{Sengupta2002}
\bibinfo{author}{\bibfnamefont{P.}~\bibnamefont{Sengupta}},
  \bibinfo{author}{\bibfnamefont{A.~W.} \bibnamefont{Sandvik}},
  \bibnamefont{and} \bibinfo{author}{\bibfnamefont{D.~K.}
  \bibnamefont{Campbell}}, \bibinfo{journal}{Phys. Rev. B}
  \textbf{\bibinfo{volume}{65}}, \bibinfo{pages}{155113}
  (\bibinfo{year}{2002}).

\bibitem[{\citenamefont{Sandvik et~al.}(2004)\citenamefont{Sandvik, Balents,
  and Campbell}}]{Sandvik2004}
\bibinfo{author}{\bibfnamefont{A.~W.} \bibnamefont{Sandvik}},
  \bibinfo{author}{\bibfnamefont{L.}~\bibnamefont{Balents}}, \bibnamefont{and}
  \bibinfo{author}{\bibfnamefont{D.~K.} \bibnamefont{Campbell}},
  \bibinfo{journal}{Phys. Rev. Lett.} \textbf{\bibinfo{volume}{92}},
  \bibinfo{pages}{236401} (\bibinfo{year}{2004}).

\bibitem[{\citenamefont{Zhang}(2004)}]{Zhang2004}
\bibinfo{author}{\bibfnamefont{Y.~Z.} \bibnamefont{Zhang}},
  \bibinfo{journal}{Phys. Rev. Lett.} \textbf{\bibinfo{volume}{92}},
  \bibinfo{pages}{246404} (\bibinfo{year}{2004}).

\bibitem[{\citenamefont{Ejima and Nishimoto}(2007)}]{Ejima2007}
\bibinfo{author}{\bibfnamefont{S.}~\bibnamefont{Ejima}} \bibnamefont{and}
  \bibinfo{author}{\bibfnamefont{S.}~\bibnamefont{Nishimoto}},
  \bibinfo{journal}{Phys. Rev. Lett.} \textbf{\bibinfo{volume}{99}},
  \bibinfo{pages}{216403} (\bibinfo{year}{2007}).

\bibitem[{\citenamefont{Tsuchiizu and Furusaki}(2002)}]{Tsuchiizu2002}
\bibinfo{author}{\bibfnamefont{M.}~\bibnamefont{Tsuchiizu}} \bibnamefont{and}
  \bibinfo{author}{\bibfnamefont{A.}~\bibnamefont{Furusaki}},
  \bibinfo{journal}{Phys. Rev. Lett.} \textbf{\bibinfo{volume}{88}},
  \bibinfo{pages}{056402} (\bibinfo{year}{2002}).

\bibitem[{\citenamefont{Tsuchiizu and Furusaki}(2004)}]{Tsuchiizu2004}
\bibinfo{author}{\bibfnamefont{M.}~\bibnamefont{Tsuchiizu}} \bibnamefont{and}
  \bibinfo{author}{\bibfnamefont{A.}~\bibnamefont{Furusaki}},
  \bibinfo{journal}{Phys. Rev. B} \textbf{\bibinfo{volume}{69}},
  \bibinfo{pages}{035103} (\bibinfo{year}{2004}).

\bibitem[{\citenamefont{Mund et~al.}(2009)\citenamefont{Mund, Legeza, and
  Noack}}]{Mund2009}
\bibinfo{author}{\bibfnamefont{C.}~\bibnamefont{Mund}},
  \bibinfo{author}{\bibfnamefont{O.}~\bibnamefont{Legeza}}, \bibnamefont{and}
  \bibinfo{author}{\bibfnamefont{R.~M.} \bibnamefont{Noack}},
  \bibinfo{journal}{Phys. Rev. B} \textbf{\bibinfo{volume}{79}},
  \bibinfo{pages}{245130} (\bibinfo{year}{2009}).

\bibitem[{\citenamefont{Liu and Wang}(2011)}]{Liu2011}
\bibinfo{author}{\bibfnamefont{G.-H.} \bibnamefont{Liu}} \bibnamefont{and}
  \bibinfo{author}{\bibfnamefont{C.-H.} \bibnamefont{Wang}},
  \bibinfo{journal}{Commun. Theor. Phys.} \textbf{\bibinfo{volume}{55}},
  \bibinfo{pages}{702} (\bibinfo{year}{2011}).

\bibitem[{\citenamefont{Carrasquilla et~al.}(2013)\citenamefont{Carrasquilla,
  Manmana, and Rigol}}]{Carrasquilla2013}
\bibinfo{author}{\bibfnamefont{J.}~\bibnamefont{Carrasquilla}},
  \bibinfo{author}{\bibfnamefont{S.~R.} \bibnamefont{Manmana}},
  \bibnamefont{and} \bibinfo{author}{\bibfnamefont{M.}~\bibnamefont{Rigol}},
  \bibinfo{journal}{Phys. Rev. A} \textbf{\bibinfo{volume}{87}},
  \bibinfo{pages}{043606} (\bibinfo{year}{2013}).

\bibitem[{\citenamefont{{\L}acki et~al.}(2014)\citenamefont{{\L}acki, Damski,
  and Zakrzewski}}]{Lacki2014}
\bibinfo{author}{\bibfnamefont{M.}~\bibnamefont{{\L}acki}},
  \bibinfo{author}{\bibfnamefont{B.}~\bibnamefont{Damski}}, \bibnamefont{and}
  \bibinfo{author}{\bibfnamefont{J.}~\bibnamefont{Zakrzewski}},
  \bibinfo{journal}{Phys. Rev. A} \textbf{\bibinfo{volume}{89}},
  \bibinfo{pages}{033625} (\bibinfo{year}{2014}).

\end{thebibliography}

\end{document}